\documentclass[showpacs,showkeys,aps,prd,amsfonts,eqsecnum,nofootinbib,
floatfix,natbib]{revtex4-1}
\usepackage[utf8]{inputenc}
\usepackage{amsmath,amssymb,epsfig,url,graphicx}
\usepackage[usenames,dvipsnames]{color}
\usepackage{slashed}
\usepackage[colorlinks,citecolor=blue]{hyperref}
\usepackage{color}

\begin{document}
\preprint{{\scriptsize CUMQ/HEP 198, HIP-2018-20/TH,HRI-RECAPP-2018-11} }
\title{Multileptonic signals of co-annihilating left-right supersymmetric
  dark matter}
  \author{Arindam Chatterjee}
\email{arindam.chatterjee@gmail.com}
\affiliation{Indian Statistical Institute, 203 B.T. Road, Kolkata-700108, India}
\author{Mariana Frank}
\email{mariana.frank@concordia.ca}
\affiliation{Department of Physics, Concordia University, 7141 Sherbrooke St.West, Montreal, QC, Canada H4B 1R6}
\author{Benjamin Fuks}
\email{fuks@lpthe.jussieu.fr}
\affiliation{Sorbonne Universit\'e, CNRS, Laboratoire de Physique Th\'eorique et Hautes \'Energies, LPTHE, F-75005 Paris, France, 
\& Institut Universitaire de France, 103 boulevard Saint-Michel, 75005 Paris, France}
\author{Katri Huitu}
\email{katri.huitu@helsinki.fi}
\affiliation{Department of Physics, and Helsinki Institute of Physics, P. O. Box 64, FI-00014 University of Helsinki, Finland}
\author{Subhadeep Mondal}
\email{subhadeep.mondal@helsinki.fi}
\affiliation{Department of Physics, and Helsinki Institute of Physics, P. O. Box 64, FI-00014 University of Helsinki, Finland}
\author{Santosh Kumar Rai}
\email{skrai@hri.res.in}
\affiliation{Regional Centre for Accelerator-based Particle Physics,
  Harish-Chandra Research Institute, HBNI, Chhatnag Road, Jhusi, Allahabad 211019, India}
\author{Harri Waltari}
\email{harri.waltari@helsinki.fi}
\affiliation{Department of Physics and Astronomy, University of Southampton, Highfield, Southampton SO17 1BJ, United Kingdom, 
\& Department of Physics, and Helsinki Institute of Physics, P. O. Box 64, FI-00014 University of Helsinki, Finland}

\date{\today}

\begin{abstract}
We perform a comprehensive dark matter analysis of left-right
  supersymmetric scenarios that includes constraints from dark matter direct
  and indirect detection experiments and that presents
  distinctive features from those available in minimal supersymmetry. We
  concentrate on dark matter candidates
  which, while satisfying all constraints, are different from those of the
  minimal supersymmetric standard model. We consider in our analysis all
  possible co-annihilation channels relevant for setups in which several states
  are light and nearly degenerate, and devise a set of representative benchmark
  points,  requiring co-annihilations, which satisfy all restrictions. We then study their consequent LHC
  signals, which exhibit promising new multileptonic signatures involving $W_R$, that 
  if observed, would provide a strong support for left-right supersymmetry.
\end{abstract}

\maketitle
\section{Introduction}
\label{sec:intro}
The LHC experiments have probed the Standard Model (SM) at high energies with no
clear signs of new physics so far. Nevertheless, it is well known that the SM
needs to be extended, as neutrino oscillation experiments show that neutrinos
have masses~\cite{Fukuda:1998mi,Aguilar:2001ty,Ahn:2002up,Abe:2011sj,An:2013zwz}
and there is convincing evidence for cold dark matter from galaxy rotation
curves~\cite{Zwicky:1933gu,Rubin:1970zza}, the cosmic microwave
background~\cite{Ade:2015xua} and the Bullet cluster observations~\cite{%
Clowe:2006eq}. In addition, the SM has features that do not have a proper
explanation, like for instance parity violation and the strong $CP$ problem. In
the SM, the Higgs boson mass term gets quadratic corrections that are
proportional to the scale of new physics, so that we expect some kind of a
cutoff mechanism to exist not too far above the electroweak scale. Supersymmetry
(SUSY) offers such a mechanism as the quadratic corrections stemming from
bosonic and fermionic states cancel, even if SUSY is softly broken.

Within the framework of SUSY, left-right symmetric (LR) models~\cite{%
Pati:1974yy,Mohapatra:1974hk} have attractive features. As they are based on the
$SU(3)_{\rm C}\times SU(2)_{\rm L}\times SU(2)_{\rm R}\times U(1)_{B-L}$ gauge
group, $R$-parity violation is forbidden as it would break $U(1)_{B-L}$. In
addition, parity breaking is dynamical and the strong $CP$ problem gets solved
as the parity violating QCD $\theta$-term is absent at tree-level and is only
generated at the two-loop level~\cite{Mohapatra:1995xd,Kuchimanchi:1995rp,%
Mohapatra:1996vg}. Moreover, LR symmetry requires the existence of right-handed
neutrinos, so that neutrinos are naturally massive, although the actual
implementation of a seesaw mechanism is not straightforward as right-handed
neutrino bare
mass terms are forbidden by the model gauge symmetry. The usual solution
requires the presence of chiral $SU(2)_{\rm R}$ triplet superfields with
non-zero $B-L$ quantum numbers. Their neutral scalar components then break
lepton number spontaneously, which generates right-handed neutrino mass terms.

Breaking parity with $SU(2)_{\rm R}$ triplets leads to a tree-level scalar
potential that violates either charge conservation or $R$-parity invariance~\cite{%
Kuchimanchi:1993jg}. The former is unacceptable and the latter
makes the lightest supersymmetric particle (LSP) unstable, so that it could
not be a viable dark matter (DM) candidate anymore. Without extending the
particle content further, the charge and $R$-parity conserving minimum can,
however, be stabilized by including one-loop corrections to the scalar
potential~\cite{Babu:2008ep,Basso:2015pka}. This both saves the LSP as a viable
DM candidate and also forces the LR symmetry breaking scale to be relatively
low, the latter yielding hopes of finding left-right supersymmetry (LRSUSY) at
the LHC.

LRSUSY has a number of viable DM candidates. The model has twelve neutralinos,
and both gaugino-dominated and higgsino-dominated states are acceptable DM
candidates. In addition, right-handed sneutrinos may annihilate efficiently
enough through gauge interactions to satisfy the relic density constraints from
Planck without the need of mixing left- and right-handed sneutrino states. In a
previous work~\cite{Frank:2017tsm}, we have analyzed right-handed sneutrinos and
gauginos LSP as candidates for dark matter, but when the LSP is much lighter
than the next-to-lightest superpartner (NLSP) so that co-annihilations can be
ignored. In this work, we relax this assumption and extend our analysis to other
possibilities where co-annihilation channels matter. As the model features two
Higgs bidoublets, there are generally several nearly-degenerate higgsino-like
neutralinos and charginos with a mass close to the effective off-diagonal Higgs
mass mixing parameter $\mu_{\mathrm{eff}}$. Co-annihilations are hence always
present and relevant both for a higgsino-like DM candidate and when the LSP is
close in mass of the higgsinos. We furthermore also examine prospects for DM
indirect detection, especially in the view of a right-handed sneutrino LSP
annihilating into right-handed neutrinos.

We finally study how these scenarios could emerge through multilepton production
in association with missing energy at the LHC, a collider signature that could
give strong support for the realization of LRSUSY in nature. In practice, we use
DM relic density constraints to fix the masses of the LSP and of the
co-annihilating neutralinos and charginos, and investigate, for a few
representative benchmark scenarios, the production and decay of a not too heavy
$SU(2)_{\rm R}$ $W_{R}$ boson into charginos and neutralinos. Such a channel is
usually linked to a sizeable branching fraction into leptons and missing energy,
so that the corresponding new physics signal
can be constrained by typical electroweakino searches. We therefore analyze the
sensitivity of a recent CMS SUSY search in the multileptons plus missing energy
channel~\cite{Sirunyan:2017lae} and then estimate the prospects of the
high-luminosity phase of the LHC.

Our work is organized as follows. We give a brief introduction to the LRSUSY
model version considered in section \ref{sec:model}. In section~\ref{%
sec:constraints} we proceed to consider existing constraints from collider
experiments and dark matter searches. We then select a number of benchmark
points representing different dark matter motivated model configurations in
section~\ref{sec:benchmarks} and then analyze the prospects of DM indirect
detection in section \ref{sec:indirect}, and of collider searches in
section~\ref{sec:collider}. We summarize our findings and conclude in
section~\ref{sec:conclusion}.

\section{The left-right supersymmetric model}
\label{sec:model}
Left-right supersymmetric models, based on the $SU(3)_{\rm C}\times SU(2)_{\rm
L}\times SU(2)_{\rm R}\times U(1)_{B-L}$ gauge symmetry, inherit all attractive
features of the left-right symmetry~\cite{Pati:1974yy,Mohapatra:1974hk}, whereas
they forbid, thanks to the gauged $B-L$ symmetry, any $R$-parity violating
operators problematic in the Minimal Supersymmetric Standard Model (MSSM).
The chiral matter in LRSUSY consist of three families of quark and lepton
superfields,
\begin{eqnarray}
  \!\!Q_L\!\!&=&\!\!\left (\begin{array}{c}
    u_L\\ d_L\end{array} \right) \sim \left ({\bf 3}, {\bf 2}, {\bf 1}, \frac13
    \right ), \quad
  Q_R\!=\!\left (\begin{array}{c}
    d_R\\-u_R \end{array} \right) \sim \left ( {\bf \bar 3}, {\bf 1}, {\bf 2}^*,
    -\frac13 \right ),\nonumber \\
  L_L&=&\left (\begin{array}{c}
    \nu_L\\ e_L\end{array}\right) \sim\left ({\bf 1}, {\bf 2}, {\bf 1}, -1
    \right), \quad
  L_R = \left (\begin{array}{c}
    e_R \\ -N \end{array}\right) \sim \left ({\bf 1}, {\bf 1}, {\bf 2}^*, 1
    \right),\nonumber
\end{eqnarray}
where the numbers in the brackets denote the representation under the
$SU(3)_{\rm C}$, $SU(2)_{\rm L}$, $SU(2)_{\rm R}$ and $U(1)_{B-L}$ gauge
factors. To ascribe a small magnitude of the neutrino masses and preserve
$R$-parity, the model superfield content includes both $SU(2)_{\rm L}$ and
$SU(2)_{\rm R}$ triplets of Higgs supermultiplets, in addition to two Higgs
bidoublets and one singlet,
\begin{eqnarray}
\label{eq:higgs-decomp}
  &&\Phi_1 = \left (\begin{array}{cc}
    \Phi^+_{11}&\Phi^0_{11}\\ \Phi_{12}^0& \Phi_{12}^-
    \end{array}\right) \sim \left ({\bf 1},{\bf 2}, {\bf 2}^*,0 \right), \qquad
  \Phi_2=\left (\begin{array}{cc}
    \Phi^+_{21}&\Phi^0_{21}\\ \Phi_{22}^0& \Phi_{22}^-
    \end{array}\right) \sim \left ({\bf 1}, {\bf 2}, {\bf 2}^*,0 \right),
  \nonumber \\
  &&\Delta_{L}=\left(\begin{array}{cc}
    \frac {1}{\sqrt{2}}\Delta_L^-&\Delta_L^0\\
    \Delta_{L}^{--}&-\frac{1}{\sqrt{2}}\Delta_L^-
    \end{array}\right) \sim ({\bf 1}, {\bf 3}, {\bf 1},-2),\qquad
  \delta_{L}  = \left(\begin{array}{cc}
    \frac {1}{\sqrt{2}}\delta_L^+&\delta_L^{++}\\
    \delta_{L}^{0}&-\frac{1}{\sqrt{2}}\delta_L^+
    \end{array}\right) \sim ({\bf 1}, {\bf 3}, {\bf 1}, 2),\nonumber \\
  &&\Delta_{R} = \left(\begin{array}{cc}
    \frac {1}{\sqrt{2}}\Delta_R^-&\Delta_R^0\\
    \Delta_{R}^{--}&-\frac{1}{\sqrt{2}}\Delta_R^-
    \end{array}\right) \sim ({\bf 1}, {\bf 1}, {\bf 3},-2),\qquad
  \delta_{R}  = \left(\begin{array}{cc}
    \frac {1}{\sqrt{2}}\delta_R^+&\delta_R^{++}\\
    \delta_{R}^{0}&-\frac{1}{\sqrt{2}}\delta_R^+
    \end{array}\right) \sim ({\bf 1}, {\bf 1}, {\bf 3}, 2), \nonumber \\
  &&  S \sim ({\bf 1}, {\bf 1}, {\bf 1}, 0) \ ,
\end{eqnarray}
where the numbers in the brackets again denote the representation under the
model gauge group. The superpotential of the model is given by
\begin{eqnarray}
\label{superpotential}
  W &=& Q_L^T {\bf Y}_{Q}^{(i)} \Phi_{i} Q_R +
        L_L^T {\bf Y}_{L}^{(i)} \Phi_{i} L_R +
        L_L^T {\bf h}_{LL} \delta_L L_L +
        L_R^T {\bf h}_{RR} \Delta_R L_R +
        \lambda_L\ S\ \mbox{Tr}\left[\Delta_L \delta_L\right] \nonumber \\
   &&+  \lambda_R\ S\ \mbox{Tr}\left[\Delta_R \delta_R\right] +
        \lambda_3\ S\ \mbox{Tr}\left[\tau_2 \Phi^T_1 \tau_2 \Phi_2\right] +
        \lambda_4\ S\ \mbox{Tr}\left[\tau_2 \Phi^T_1 \tau_2 \Phi_1\right]
   \nonumber \\
   &&+  \lambda_5\ S\ \mbox{Tr}\left[\tau_2 \Phi^T_2 \tau_2 \Phi_2\right] +
        \lambda_S\ S^3 + \xi_F\ S \ ,
\end{eqnarray}
where the Yukawa couplings ${\bf Y}_{Q,L}$ and ${\bf h}_{LL,RR}$ are $3\times3$
matrices in flavor space, the $\lambda$ parameters stand for the strengths of
the various Higgs(ino) interactions, with $\tau_2$ being the second Pauli
matrix, and $\xi$ consists in a linear singlet term. The Lagrangian of the model
includes, on top of usual SUSY gauge interaction, kinetic and superpotential
interaction terms, soft SUSY-breaking terms (standard scalar and gaugino mass
terms together with multi-scalar interactions deduced from the form of the
superpotential). Explicit expressions and more details about the model can be
found in Refs.~\cite{Alloul:2013fra, Frank:2017tsm}.

The neutral component of the $SU(2)_{\rm R}$ Higgs scalar field $\Delta_R$
acquires a large vacuum expectation value (VEV) $v_{R}$, which breaks the LR
symmetry and makes the $SU(2)_{\rm R}$ gauge sector heavy. The complete set of
non-vanishing VEVs responsible for breaking the symmetry down to $U(1)_{\rm em}$
reads
\begin{eqnarray} \label{eq:higgs-vevs}
  &&\langle \Phi_1 \rangle = \left (\begin{array}{cc}
    0 & 0\\ \frac{v_1}{\sqrt{2}} &0
    \end{array}\right), \quad
  \langle \Phi_2 \rangle = \left (\begin{array}{cc}
    0 & \frac{v_2}{\sqrt{2}}\\ 0  & 0
    \end{array}\right) , \quad
  \langle\Delta_{R}\rangle = \left(\begin{array}{cc}
    0&\frac{v_R}{\sqrt{2}} \\
    0&0 \end{array}\right),\quad
  \langle \delta_{R}\rangle  = \left(\begin{array}{cc}
    0&0\\
    \frac{v_R^\prime}{\sqrt{2}} &0 \end{array}\right), \nonumber\\
  &&\langle S \rangle =\frac{v_S}{\sqrt{2}} \,.
\end{eqnarray}
This vacuum structure allows for avoiding constraints from electroweak precision
tests, flavor-changing neutral currents and a too large mixing between the $W_L$
and $W_{R}$ bosons. It is moreover stable provided that $\lambda_{4}=\lambda_{5}=0$. In
order to prevent the tree-level vacuum state from being a charge-breaking one,
one can either rely on spontaneous $R$-parity violation~\cite{%
Kuchimanchi:1993jg}, one-loop corrections~\cite{Babu:2008ep,Basso:2015pka},
higher-dimensional operators~\cite{Mohapatra:1995xd} or additional $B-L=0$
triplets~\cite{Aulakh:1997ba}. Whereas the first two options restrict $v_{R}$ to
be of at most about $10$~TeV, the latter ones enforce $v_{R}$ to lie above
$10^{10}$~GeV. In this work, we rely on radiative corrections to stabilize the
vacuum, so that the LSP is stable and can act
as a DM candidate~\cite{Frank:2017tsm}. Two viable LSP options emerge from
LRSUSY, neutralinos and right sneutrinos.

\subsection{Neutralinos and charginos}

The model has twelve neutralinos whose mass matrix can be decomposed into three
independent blocks, two $2\times 2$ blocks describing the mixing of the
$\tilde{\delta}_L$/$\tilde{\Delta}_L$ and
$\tilde{\Phi}_{22}^0$/$\tilde{\Phi}_{11}^0$ fields respectively, and one
$8\times 8$ block related to the mixing of the eight other neutral
higgsinos and gauginos.  The two dimension-two blocks
are given, in the $(\tilde{\delta}_L^0, \tilde{\Delta}_L^0)$ and
$(\tilde{\Phi}_{22}^0, \tilde{\Phi}_{11}^0)$ bases, by
\begin{equation}
M_{\tilde{\chi}_{\delta}}=
\begin{pmatrix}
0 & \mu_{L}\\
\mu_{L} & 0
\end{pmatrix}\qquad\text{and}\qquad
M_{\tilde{\chi}_{\Phi}}=
\begin{pmatrix}
0 & -\mu_{\mathrm{eff}}\\
-\mu_{\mathrm{eff}} & 0
\end{pmatrix}\ ,
\end{equation}
while the last $8\times 8$ block reads, in the $(\tilde{\Phi}_{12}^0, \tilde{\Phi}_{21}^0,
\tilde{\delta}_{R}^0, \tilde{\Delta}_{R}^0, \tilde{S}, \tilde{B}, \tilde{W}_{L}^{0},
\tilde{W}_{R}^{0})$ basis,
\begin{equation}
M_{\tilde{\chi}^{0}}  = 
\begin{pmatrix}
0 & -\mu_{\mathrm{eff}} & 0 & 0 & -\mu_{d} & 0 & \frac{g_{L}v_{u}}{\sqrt{2}} & -\frac{g_{R}v_{u}}{\sqrt{2}} \\
- \mu_{\mathrm{eff}} & 0 & 0 & 0 & -\mu_{u} & 0 & -\frac{g_{L}v_{d}}{\sqrt{2}} & \frac{g_{R}v_{d}}{\sqrt{2}}\\
0 & 0 & 0 & \mu_{R} & \frac{\lambda_{R} v_{R}^{\prime}}{\sqrt{2}} & g^\prime v_{R} & 0 & -g_{R}v_{R}\\
0 & 0 & \mu_{R} & 0 & \frac{\lambda_{R} v_{R}}{\sqrt{2}} & -g^\prime {v}_{R}^{\prime} & 0 & -g_{R}{v}_{R}^{\prime}\\
-\mu_{d} & -\mu_{u} & \frac{\lambda_{R} {v}_{R}^{\prime}}{\sqrt{2}} & \frac{\lambda_{R} v_{R}}{\sqrt{2}} & \mu_S & 0 & 0 & 0\\
0 & 0 & g'v_{R} & -g^\prime{v}_{R}^{\prime} & 0 & M_{1} & 0 & 0\\
\frac{g_{L}v_{u}}{\sqrt{2}} & -\frac{g_{L}v_{d}}{\sqrt{2}} & 0 & 0 & 0 & 0 & M_{2L} & 0\\
-\frac{g_{R}v_{u}}{\sqrt{2}} & \frac{g_{R}v_{d}}{\sqrt{2}} & -g_{R}v_{R} & -g_{R}{v}_{R}^{\prime} & 0 & 0 & 0 & M_{2R}
\end{pmatrix}\ .
\label{eq:Xino8}\end{equation}
In the above expressions, we have defined $\mu_{\mathrm{eff}}=\lambda_{3}\frac{v_{S}}{\sqrt{2}}$, $\mu_S=\lambda_S \frac{v_S}{\sqrt{2}}$, $\mu_{L,R}=\lambda_{L,R}\frac{v_{S}}{\sqrt{2}}$ and $\mu_{u,d}=\lambda_{3} \frac{v_{u,d}}{\sqrt{2}}\,$.

We assume that the $\lambda_{R}$ coupling is large, as this helps to stabilize
the vacuum state~\cite{Basso:2015pka}. Invoking LR symmetry, $\lambda_L$ is
taken large too. After accounting for the fact that $v_{R}$, $v_{R}^{\prime}$ (responsible for LR breaking) and
$v_S$ (induced by SUSY breaking) are large as well, it turns out that many of the neutralino states are
heavy. Gauginos, whose masses originate in contrast from soft terms, can be
light (with the exception of the $\tilde{W}_{R}$ wino that has a mass close to the
$W_{R}$ boson mass). 
The LR bidoublet higgsinos, which are nearly degenerate with masses close
to $|\mu_{\mathrm{eff}}|$, can also be light. As shown by the non-zero elements of the mass matrix, Eq.~\eqref{eq:Xino8}, two of the higgsinos mix with the $\tilde{W}_{L,R}$
winos with a strength depending on $\tan\beta$. Typically then, higgsino-dominated
states  acquire a gaugino component that can be up to $10\%$.
Due to the breaking of the $SU(2)_{\mathrm{R}}\times U(1)_{B-L}$
symmetry into $U(1)_{\mathrm{Y}}$, the $\tilde{B}$ and $\tilde{W}_{R}$ gauginos
mix. If one of these admixtures is close in mass to the bidoublet higgsinos,
further mixings occur, as above-mentioned, and the degeneracy of the higgsinos
states is lifted. In this case, the gaugino-dominated state becomes further split from
the rest of the neutralino spectrum.

In the charged sector, the model has six singly-charged
charginos whose mass matrix is written, in the 
($\tilde{\delta}_{L}^+$,
$\tilde{\delta}_{R}^+$, $\tilde{\Phi}_{11}^+$, $\tilde{\Phi}_{21}^+$, $\tilde{W}_L^+$,
$\tilde{W}_{R}^+$) and $(\tilde{\Delta}_{L}^-, \tilde{\Delta}_{R}^-,
\tilde{\Phi}_{12}^-, \tilde{\Phi}_{22}^-, \tilde{W}_L^-, \tilde{W}_{R}^-)$ bases
as
\begin{equation}
\label{eq:mchargino}
M_{\tilde{\chi}^{\pm}}=
\begin{pmatrix}
\lambda_{L}v_{S}/\sqrt{2} & 0 & 0 & 0 & 0 & 0\\
0 & \lambda_{R} v_{S}/\sqrt{2} & 0 & 0 & 0 & -g_{R}v_{R}\\
0 & 0 & 0 & \mu_{\mathrm{eff}} & g_{L}v_{u}/\sqrt{2} & 0\\
0 & 0 & \mu_{\mathrm{eff}} & 0 & 0 & -g_{R}v_{d}/\sqrt{2}\\
0 & 0 & 0 & g_{L}v_{d}/\sqrt{2} & M_{2L} & 0\\
0 & g_{R}v_{R}^{\prime} & -g_{R}v_{u}/\sqrt{2} & 0 & 0 & M_{2R}\\
\end{pmatrix}.
\end{equation}
Here again the $\tilde{\Delta}_{L,R}^{\pm}$ states are heavy and the bidoublet
charginos have a mass close to $|\mu_{\mathrm{eff}}|$, so that they will be
nearly degenerate with the corresponding neutralinos. The same is true for the
$SU(2)_{\rm L}$ winos whose charged and neutral states are almost degenerate,
but the situation is different for the $SU(2)_{\rm R}$ sector where the more
complex mixing pattern in the neutralino sector lifts all potential degeneracy.

\subsection{Dark matter candidates}
\label{subsec:dmc}

Of the twelve neutralinos of the model, gaugino-like and LR
bidoublet higgsino-like neutralinos can generally be lighter than 1~TeV.
The
correct relic density can, however, only be accommodated with dominantly-bino-like
LSP with a mass close to $m_{h}/2$~\cite{Frank:2017tsm}, whilst in the bidoublet
higgsino case (featuring four neutralinos and two charginos that are
nearly-degenerate), co-annihilations play a crucial role and impose higgsino
masses close to 700~GeV. Hence the higgsino LSP case is an example of a heavy
and compressed spectrum, which poses a challenge for direct searches for SUSY.

Right sneutrino LSPs annihilate via the exchange of an $s$-channel Higgs boson
through gauge interactions stemming from the $D$-terms~\cite{Frank:2017tsm}.
Without options for co-annihilating, the LSP sneutrino mass must lie between
$250$ and $300$~GeV, heavier masses leading to DM overproduction. However,
potential co-annihilations with neutralinos enhance the effective annihilation
cross section so that the relic density constraints can be satisfied with
heavier sneutrinos. The fully degenerate sneutrino/higgsino scenario imposes an
upper limit on the LSP sneutrino mass of $700$~GeV. Additionally, together with
the LSP, right neutrinos can also be part of the dark sector~\cite{%
Bhattacharya:2013nya}.

As direct detection constraints imposed by the XENON1T~\cite{Aprile:2018dbl} and
PANDA \cite{Cui:2017nnn} collaborations have put light DM scenarios under severe
scrutiny, viable LRSUSY DM setups accounting for the relic density and direct
detection constraints simultaneously need to rely on various co-annihilation
options. In this work we consider several of such scenarios with different LSP
options, and additionally highlight the corresponding implications for searches
at the LHC.

A robust signal of left-right symmetry consists in the discovery of an
$SU(2)_{\rm R}$ gauge boson $W_R$, possibly together with a right neutrino
$N$. From a pure spectral analysis, the SUSY nature of such a signal could
originate from the above dark matter considerations that lead to favored LRSUSY
scenarios in which several neutralinos and charginos are light. This hence
motivates LRSUSY investigations through a new $W_R$ boson search channel, where
decays into pairs of electroweakinos are considered, the corresponding branching
ratio being as large as 25\% in many LRSUSY scenarios. Especially when a
sneutrino is the LSP, we expect that the decays of these neutralinos and
charginos lead to leptonic final states at colliders, so that the production of
multileptonic systems in association with a large amount of missing transverse
energy $\slashed{E}_T$ is enhanced. Whilst such a multilepton signal with
missing energy is a characteristic SUSY collider signal, it also provides an
additional search channel for $W_R$ bosons. The resonant production
mode offers the opportunity to reconstruct the $W_R$ boson mass through the
study of kinematic thresholds featured by various transverse observables.

Before proceeding to the analysis of promising collider signals, we review in
the next section the pertinent features and constraints imposed on the
parameters of the model, including those coming from dark matter.
\vskip0.2in
\section{Constraints on the spectrum}
\label{sec:constraints}
\subsection{The Higgs sector}
\label{subsec:higgs}

The considered LRSUSY version has a relatively light $SU(2)_{\rm R}$
doubly-charged
Higgs boson with a mass originating from loop corrections~\cite{Babu:2008ep,%
Frank:2011jia,Babu:2014vba,Basso:2015pka}. Whereas the ATLAS collaboration has
excluded doubly-charged Higgs masses ranging up to $650$--$760$~GeV when the
doubly-charged Higgs boson decays exclusively into same-sign electrons or
muons~\cite{Aaboud:2017qph}, masses of about 300~GeV are still allowed when the
branching ratio in these modes is of at most a few percents.
In contrast, the CMS collaboration has searched for doubly-charged Higgs bosons
in all leptonic channels but interpreted the results only in the $SU(2)_{\rm L}$
case. As the associated production of an $SU(2)_{\rm R}$ doubly-charged Higgs
boson with a singly-charged one is suppressed by the $W_R$ mass and as any
neutral current production mode is weaker by virtue of reduced couplings,
the limit of $396$~GeV~\cite{CMS:2017pet} can be reduced to about
$300$--$350$~GeV, depending on the branching ratio into tau pairs.

Although one could enforce the doubly-charged Higgs boson to be heavy enough to
be compliant with all current bounds, we prefer imposing that it decays mainly
into tau leptons. We fix the different branching ratios to
\begin{equation}
  {\rm BR}(H^{\pm\pm}\rightarrow \tau^{\pm}\tau^{\pm}) = 92\%\ , \qquad
  {\rm BR}(H^{\pm\pm}\rightarrow \mu^{\pm} \mu^{\pm})  =
  {\rm BR}(H^{\pm\pm}\rightarrow  e^{\pm}   e^{\pm})   = 4\%\ ,
\end{equation}
so that the mass of the $SU(2)_{\rm R}$ doubly-charged Higgs boson can be safely
set to about $350$~GeV, the exact value being not crucial for our discussion.
The important parameter consists instead in the Yukawa texture, which also
determines right-handed neutrino masses and contributes to the sneutrino mass
matrix.

We impose that the SM-like Higgs boson mass $m_h$ is compatible with \cite{Aad:2012tfa,Chatrchyan:2012xdj}
\begin{equation}
  m_h = 125.1\pm 0.3~{\rm GeV}\ ,
\end{equation}
the uncertainty being chosen smaller than the corresponding theoretical
error. The Higgs mass is an essential input in the relic density computation,
due to $s$-channel Higgs boson exchange contributions, so that we want that
input to be reasonably close to the experimentally-measured value. The SM-like
Higgs boson mass is mostly affected by $\tan\beta$ and the stop masses and
mixings. As the tree-level mass is larger than in the MSSM~\cite{Babu:1987kp,%
Huitu:1997rr}, a $125$~GeV mass value can always be achieved with rather
moderate stop masses and mixings. However, $\tan \beta$
cannot have  too small a value as the tree-level Higgs mass vanishes in the
$\tan\beta\rightarrow 1$ limit, like in the MSSM. With $\tan\beta\gtrsim 5$, we
obtain a SM-like Higgs boson mass compatible with the experimental value, for stop
masses of a couple of TeV.

The second $CP$-even state, the lightest $CP$-odd state and the lightest charged
Higgs bosons are predicted to be nearly degenerate and of mass of about $m_A$.
The most stringent constraint on $m_A$ comes from the $B_{s}\rightarrow \mu\mu$
decay, to which the $CP$-odd state can yield a sizeable contribution, 
which is enhanced for large values of $\tan\beta$~\cite{%
Babu:1999hn}. We therefore use moderate values for $\tan\beta$. Starting from
\begin{equation}
  m_{A}^{2}\sim g_{R}^{2}v_{R}^{2}(\tan^{2}\beta_{R}-1)\ ,
\end{equation}
we correlate $v_{R}$ and $\tan\beta_{R}$ so that the resulting masses are close
to $650$~GeV, a value that is in addition compatible with direct search results.
In those notations, $\tan\beta_R$ stands for the ratio of the two
$SU(2)_{\rm R}$ Higgs triplet VEVs and $g_R$ for the $SU(2)_{\rm R}$ coupling
constant. On the other hand, if $\tan\beta_{R}$ deviates too
much from 1, it gives a negative contribution to the doubly-charged Higgs-boson
mass, thus we impose values close to $1.05$ for $\tan\beta_{R}$.

All other Higgs bosons have masses of the scale of either $v_{R}$ or $v_{S}$,
{\it i.e.} of several TeV, so they do not yield any constraint on the model.

\subsection{Right-handed neutrinos}
Generic searches for right-handed neutrinos were performed at LEP~\cite{%
Achard:2001qw}, leading to bounds on right-handed Majorana neutrino masses of at
most $90.7$~GeV. In our model the right-handed neutrino mass matrix reads
\begin{equation}
({\bf m_{N}})_{ij}= ({\bf h}_{RR})_{ij}\ v_{R}\ ,
\end{equation}
where ${\bf h}_{RR}$ also dictates the different doubly-charged Higgs branching
ratio. As we have enforced the $SU(2)_{\rm R}$ doubly-charged Higgs boson to
decay mainly into taus, the right-handed tau neutrino turns out to be
significantly heavier than the others. With our choices of $v_{R}$ (essentially
determined by the bounds on the $W_{R}$ boson mass as shown in
section~\ref{subsec:WR}), the electron and muon right-handed neutrino masses
are close to $150$~GeV, whereas the right-handed tau neutrino mass is of about
$750$~GeV.

Such a spectrum implies that $t$-channel neutralino-mediated sneutrino DM
annihilation into right-handed neutrinos is kinematically open only for electron
or muon
sneutrinos, but not for the tau ones, once the mass constraints originating from
the DM relic density are accounted for (see section~\ref{subsec:darkmatter}).

\subsection{The $SU(2)_{\rm R}$ gauge sector}
\label{subsec:WR}
In LRSUSY, the masses of the $W_{R}$ and $Z_R$ bosons are related and the $Z_R$
is always heavier than the $W_{R}$ boson. Hence $W_{R}$ searches are more
restrictive. The $W_{R}$ boson can decay into jets and, if the $SU(2)_{\rm R}$
gauge coupling equals the $SU(2)_{\rm L}$ one ($g_L=g_R$), limits on sequential
$W^{\prime}$ bosons can be reinterpreted straightforwardly. ATLAS and CMS have
obtained bounds of $3.6$~TeV~\cite{Aaboud:2017yvp} and $3.3$~TeV~\cite{%
Sirunyan:2018xlo}, on such a sequential extra gauge boson, respectively. After
taking into account the 20--25\% chance that the $W_{R}$ boson decays into
superpartners, bounds turn to be slightly weaker. We conservatively adopt, in
our analysis, a $W_{R}$ mass,
\begin{equation}
  m_{W_R} \gtrsim 3.3~{\rm TeV} \ ,
\end{equation}
that allows for evading those constraints. For such values of 
$W_R$-boson masses, the corresponding neutral $Z_{R}$ boson is heavier, with a
mass of about $5.6$~TeV. This is compatible with the current experimental lower
bounds for $Z^{\prime}$ bosons decaying into lepton pairs, which restrict $Z^\prime$ mass 
to be slightly heavier than about $4$~TeV~\cite{Aaboud:2017buh,%
Sirunyan:2018exx}.

The other decay mode which has been heavily investigated at the LHC consists in
decays into an associated $\ell N_{\ell}$ pair which gives rise to an
$\ell\ell jj$ signature. The Majorana nature of the right-handed neutrino allows
for probing both the same-sign and opposite-sign dilepton channels~\cite{%
Keung:1983uu}. Both the ATLAS and CMS collaborations have looked for such a
$W_R$ signal, excluding $W_R$ masses up to about 4.7~TeV for right-handed
(muon or electron) neutrino masses lying between $500$~GeV and $2$~TeV~\cite{%
Sirunyan:2018pom,Aaboud:2018spl}. For lower right-handed neutrino masses 
below $200$~GeV, which corresponds to our case, the bound is close to $3$~TeV 
(taking into account the suppression from SUSY decay modes) and hence less restrictive than 
the one originating from dijet searches. For the tau channel, it is even a lot weaker, with
$m_{W_R}$ excluded at most at $2.9$~TeV~\cite{Sirunyan:2017yrk},
again without assuming decays to superpartners.

\subsection{Dark matter constraints}
\label{subsec:darkmatter}

\subsubsection{The no co-annihilation case}
\label{subsec:noco}
\begin{figure}
\begin{center}
\includegraphics[width=6.6cm,height=6.6cm]{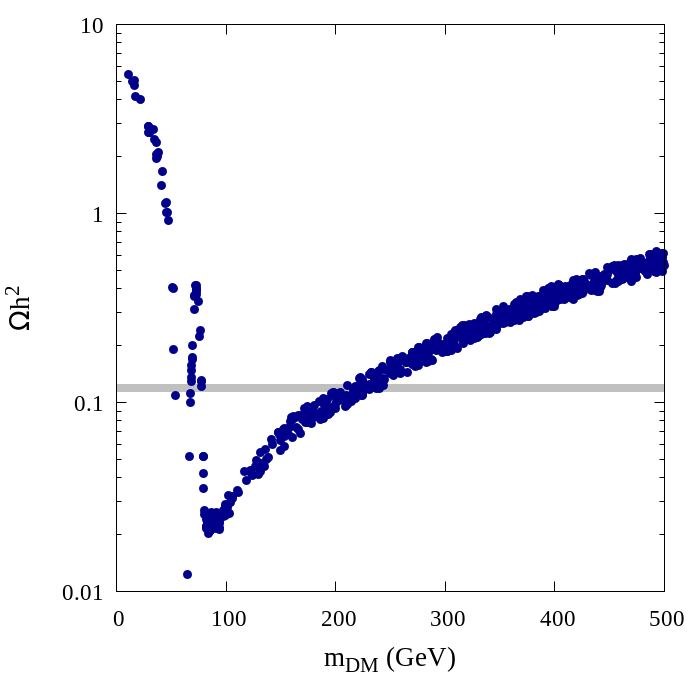}
\includegraphics[width=6.6cm,height=6.6cm]{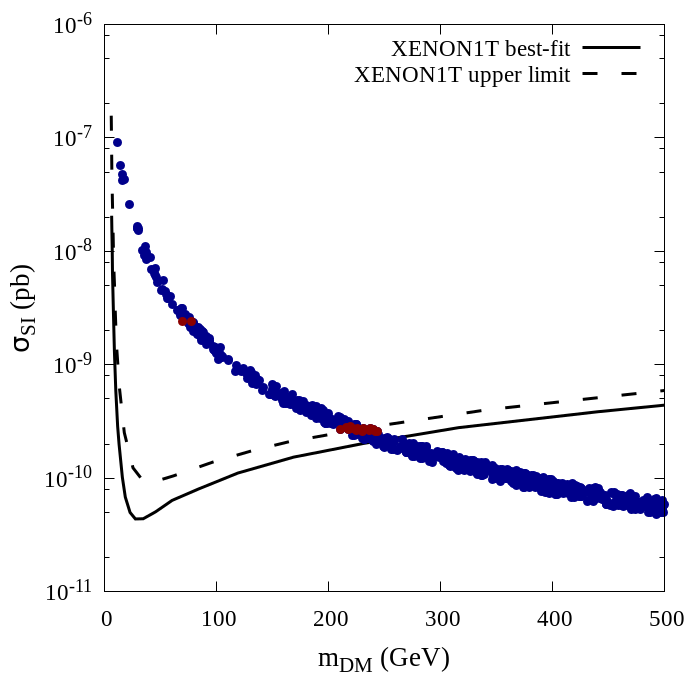}
\end{center}
\caption{Dependence of the relic density (left) and direct detection cross
  section (right) on the right-handed sneutrino mass. We find a viable solution,
  with respect to present cosmological data, for sneutrino masses around
  250~GeV. The results do not include DM co-annihilation with any other
  sparticle.} \label{fig:snu_dm}
\end{figure}

Sneutrino DM mostly annihilates, in LRSUSY scenarios, through an $s$-channel
exchange of the 125~GeV Higgs boson $h$. This contrasts with models in which
the right-handed sneutrino is a singlet. In this latter case, sneutrino
annihilations have indeed to rely either on resonances to reproduce the observed
relic abundance~\cite{BhupalDev:2012ru,DeRomeri:2012qd,Banerjee:2013fga,%
Chang:2018agk}, or on the mixing of left and right sneutrinos~\cite{%
ArkaniHamed:2000bq,Dumont:2012ee,Chatterjee:2014bva,Arina:2015uea}. In LRSUSY
models, right-handed sneutrinos are part of right-handed sleptonic doublets,
so that their coupling to the Higgs boson $h$ reads
\begin{equation}\label{eq:RHgauge}
\lambda_{h\widetilde\nu_R\widetilde\nu_R} = \frac{1}{4}g_R^2\,v \sin(\alpha + \beta)\ ,
\end{equation}
where $g_{R}$ is the $SU(2)_{\rm R}$ gauge coupling, $v=246$~GeV is the SM Higgs
VEV, $\alpha$ stands for the mixing angle of the $CP$-even Higgs states and
$\tan \beta=v_2/v_1$. We performed a scan to ascertain the sneutrino
mass regions that can produce correct relic density, and present the results in
Figure~\ref{fig:snu_dm}. Our DM relic calculation is performed with
\textsc{MadDM 2.0} \cite{Backovic:2015cra}, and our scan procedure deliberately
omits any potential co-annihilation channel (see section~\ref{subsec:sneutrinos}
for the impact of the co-annihilation channels). As 
detailed below, the particle spectrum and the necessary UFO
version~\cite{Degrande:2011ua} of the model have been generated with
{\sc Sarah}~4~\cite{Staub:2013tta,Basso:2015pka} and \textsc{SPheno 3}~\cite{Porod:2011nf}.

The results demonstrate that the structure of the right-handed sneutrino gauge
coupling of Eq.~\eqref{eq:RHgauge} alone can lead to sufficient annihilations,
even further away from the Higgs-funnel resonant region. The horizontal grey
band on the left panel of the figure indicates the $2\sigma$
experimentally-allowed range as derived by the Planck
collaboration~\cite{Ade:2015xua},
\begin{equation}
  \Omega h^2 \in [0.1163, 0.1217 ] \ ,
\label{eq:Planck}\end{equation}
whilst on the right panel, we report the distribution of the direct detection
cross section, $\sigma_{SI}$, as a function of the sneutrino mass. In this
subfigure, we mark as red points those points consistent with the measured relic
abundance (as obtained on the left figure). Those points, however, turn out to be
ruled out by the best-fit exclusion provided by the XENON1T
collaboration~\cite{Aprile:2018dbl}, shown as the solid black line in the
figure. We could alternatively rely on the XENON1T $2\sigma$ upper limit on
$\sigma_{SI}$ presented as a dashed black line, given the large uncertainties
that contribute to the cross-section measurements. This leads to the existence
of viable configurations that would merit further attention. However, the
expected progress in future DM direct detection experiments will challenge these
benchmarks, and could potentially exclude the
full hypothesis of sneutrino DM in left-right supersymmetry. Direct detection
would indeed push for heavier sneutrinos, which turn out to yield over-abundant
DM. Including co-annihilations may, however, modify these conclusions.

\subsubsection{Co-annihilations}
\label{subsec:sneutrinos}
If there are superpartners that are close in mass to the LSP, they are present
when dark matter freezes out and co-annihilation processes need to be taken into
account~\cite{Edsjo:1997bg}. Charginos and neutralinos
annihilate more efficiently to SM particles than sneutrinos. 
Co-annihilations consequently reduce the relic density
relative to the no-co-annihilation case, although the effect is
Boltzmann-suppressed when the mass difference between the LSP and the
co-annihilating particles becomes larger. In this work, we mainly focus on
co-annihilations of the sneutrino with NLSP neutralino and/or chargino states,
LRSUSY
models having altogether twelve neutralinos and six singly-charged charginos.
Whilst most states are naturally in the multi-TeV range, some may be lighter and
thus relevant from a cosmological standpoint. Their masses are controlled by the
soft supersymmetry-breaking parameters for what concerns the gauginos, while the
higgsinos have a mass of the order of $\mu_{\rm eff}$. Light higgsinos consist
in an appealing option, as there are four neutral and two charged
nearly-degenerate bidoublet higgsinos that could potentially yield sizeable
effects on the relic density.

The bidoublet higgsinos form a nearly degenerate set of four neutralinos and
two charginos and hence co-annihilations are always present if the lightest of
these neutralinos is either the LSP, or the NLSP in the case where it is nearly
degenerate with the LSP.
The higgsinos co-annihilate mainly via the $\tilde\chi_i^0\tilde\chi_j^\pm\to
q\overline{q}'$ and $\tilde{\chi}_{i}^{0}\tilde{\chi}_{j}^{0}\to q\overline{q}$
or $VV$ ($V=W,Z$) channels, processes that are all  mediated mainly by $s$-channel $W$ boson,
$Z$ boson, and Higgs boson exchanges with the mediator depending on
the charges and $CP$ properties of the co-annihilating particles. Annihilations
into quarks via gauge boson exchanges are often the dominant channels
and the relevant couplings  here are standard electroweak gauge couplings.

If  a sneutrino LSP  is mostly degenerate with the higgsinos,
additional co-annihilations with the sneutrino need to be considered. The most
significant of these modes consists in $\tilde{\nu}\tilde{\chi}^{0}\to\ell^{\pm}
W^{\mp}$ co-annihilations, which proceed via a $t$-channel wino exchange. Since
this channel requires either a mixing between the left- and right-handed
sneutrinos, or between the left- and right-handed charged winos, both mixings
that are small in our model, the corresponding contributions to the relic
abundance are relatively small compared to the $\tilde\nu\tilde\nu\to VV$ or $t\bar t$
modes. For cases in which the splittings between the sneutrino LSP and the
lighter neutralinos and charginos are small, it however turns out that neutralino-pair and
neutralino-chargino annihilation cross sections are one order of magnitude
larger than the sneutrino-sneutrino one (provided the Boltzmann suppression is
not too important).

One of the benchmark scenarios that will be adopted below yields a relic density
that is compatible with Planck data by involving the co-annihilations of a
sneutrino LSP with a left-handed wino NLSP. In such a case, the most important
(co-)annihilation channels consist of $\tilde\chi^0_1\tilde\chi^0_1\to W^+W^-$
and $\tilde{\chi}^{0}_{1}\tilde{\chi}^{\pm}_{1}\rightarrow ZW^{\pm}$ scattering, 
 mediated by $s$-channel $Z$-boson and $W$-boson exchanges,
respectively. As the left-handed wino does not usually mix with any other
neutralino or chargino, the effective cross section is then entirely determined by
gauge coupling strengths and the Boltzmann suppression stemming from mass
difference between the LSP and the NLSP.

\begin{figure}
\begin{center}
\includegraphics[width=0.8\textwidth]{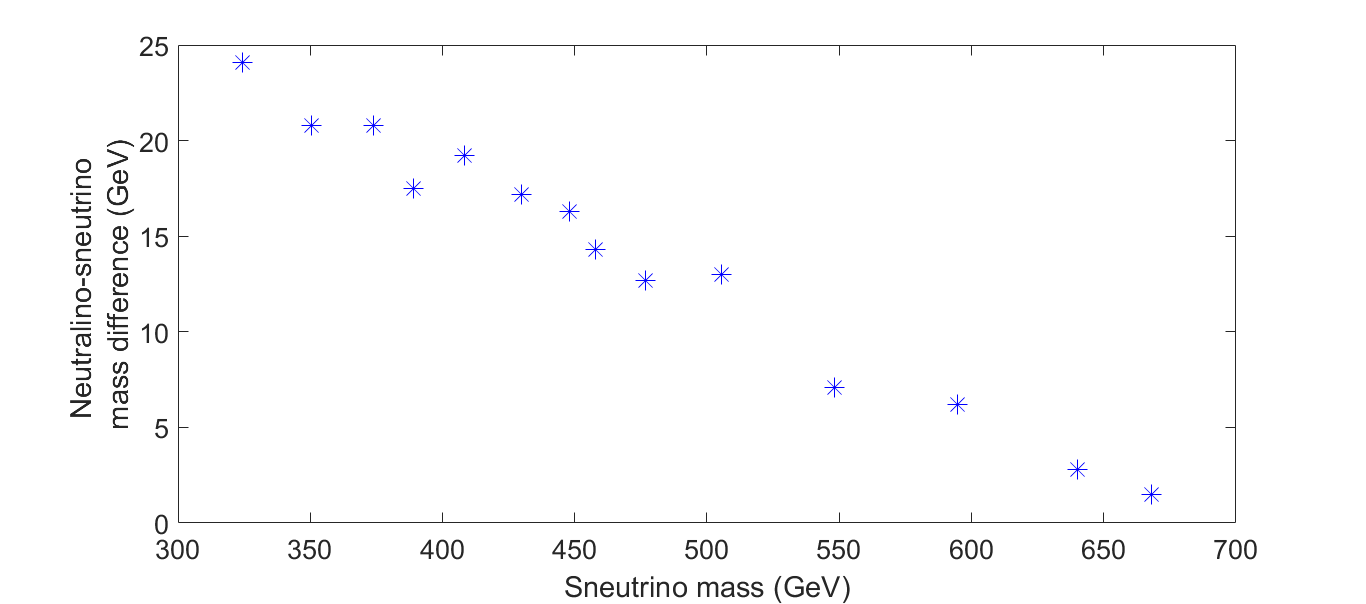}
\end{center}
\caption{Distribution of the mass difference between the sneutrino LSP and the
  NLSP, given as a function of the sneutrino mass. For each point, the relic
  density matches the Planck value thanks to the bidoublet higgsino-like
  neutralinos and charginos that are nearly degenerate and not too much heavier.
} \label{fig:snuco-annihilation}
\end{figure}

The effect of the co-annihilations on a sneutrino LSP density is illustrated in
Figure~\ref{fig:snuco-annihilation}, whilst we refer to the next subsection for
the higgsino LSP case. In Figure~\ref{fig:snuco-annihilation}, we present the
dependence of the mass difference between the sneutrino LSP and the NLSP on the
sneutrino mass. Each point corresponds to a scenario where the Planck value for
the relic density is reproduced as in Eq.~\eqref{eq:Planck}. We observe that
cosmologically-viable configurations can be found for mostly any sneutrino mass
ranging up to 675~GeV, the mass value at which the sneutrino cannot be the LSP
anymore. Comparing with the results of section~\ref{subsec:noco}, the LSP mass
can hence be viably shifted by up to several hundreds of GeV by the sole virtue
of the co-annihilation channels.

For many scanned configurations, the co-annihilating new degrees of freedom
annihilate less efficiently than the LSP. Their net effect
is a reduction of the full annihilation rate at freeze-out instead of
an enhancement, so that the relic density is increased~\cite{Profumo:2006bx}.
In LRSUSY setups with a sneutrino LSP, this happens either when some of the
heavier Higgs bosons are lighter than the LSP, or when the spectrum features
nearly degenerate sneutrinos. In addition to the SM-like Higgs boson $h$, LRSUSY
spectra indeed always feature MSSM-like Higgs states (namely a $CP$-even Higgs
boson $H$, a $CP$-odd Higgs boson $A$ and a charged Higgs boson $H^\pm$) which
are nearly degenerate. These can in
principle be lighter than the LSP and then impact the relic density in the
sneutrino LSP case through $D$-term four-point couplings that drive the
$\tilde{\nu}_{R}\tilde{\nu}_{R}\rightarrow HH$, $AA$ and $H^{+}H^{-}$
annihilation channels. Whilst such an option allows for very light sneutrino LSP
solutions with respect to the relic abundance, one cannot get a scenario where
constraints from direct searches for heavier scalars and flavor physics can
simultaneously be satisfied. A light $CP$-odd state indeed sizeably contributes to
$B_{s}\rightarrow \mu^{+}\mu^{-}$~\cite{Babu:1999hn}, which is excluded in the
light of current data. Spectra exhibiting several light and degenerate
sneutrinos are however not affected by those considerations, so that one may
push the sneutrino mass down to about $200$~GeV. This possibility is, however,
ruled out by DM direct detection bounds from XENON1T, as shown in
Figure~\ref{fig:snu_dm}.

The viable parameter space for other types of co-annihilating scenarios (featuring light $SU(2)_{\rm R}$
wino-like or higgsino-like electroweakinos for instance) are excluded by
collider searches for extra gauge bosons or doubly-charged Higgs bosons.

\subsubsection{Higgsino LSP}
\label{subsec:higgsino}
In the previous section, we focused on sneutrino LSP scenarios where
co-annihilations with nearly-degenerate bidoublet higgsinos were yielding the
observed DM relic abundance. Conversely, similar cosmologically viable setups
could be obtained when the LSP is a neutral higgsino. In this case, the relic
density increases with the LSP mass once all annihilation channels are
kinematically open. The Planck value is matched for LSP masses of around
750~GeV when co-annihilations with sneutrinos are ignored. The impact of the
latter decreases the effective annihilation cross section and then points towards
a slightly lighter LSP of about 675--700~GeV. This leads to viable spectra that
are fairly heavy, with all lighter states being
mostly degenerate bidoublet higgsinos and sneutrinos. Such a configuration would
also be roughly compatible with the AMS-02 results \cite{Aguilar:2016kjl}, which points to a TeV-scale
DM candidate.

Scenarios in which a gaugino state is nearly degenerate with
the higgsinos do not yield much differences. If the LSP is higgsino-dominated, 
with an up to 30\% gaugino admixture, the
relic density constraint can be satisfied with slightly lighter LSP masses, and
the annihilation channels are nearly the same as in a scenario where only
higgsinos would be co-annihilating. Bino-higgsino
co-annihilations, that could be crucial in the MSSM, do not work in the same way
in LRSUSY models. The bino in our model always mixes strongly with
the $SU(2)_{\rm R}$ wino, so that there is no pure bino-state at all. Therefore, if we try
to design a scenario in which a bino-state would be degenerate with the
higgsinos, the higgsinos will also mix with this bino-wino combination.
Basically we will end with two states, both admixtures of gauginos and
higgsinos. The mixing lifts the degeneracy among the higgsinos, so that one
state will be lighter and the other heavier than the original with degenerate higgsinos. This difference
with the MSSM is thus completely expected, as the MSSM ($U(1)_{\rm Y}$) bino is
here made of an admixture of the neutral $SU(2)_{\rm R}$ gaugino and the
$U(1)_{B-L}$ bino. The well-tempered MSSM scenario consists thus of a triple
admixture of states.

However, if the most gaugino-dominated state is the
LSP, the mass difference between the co-annihilating particles is larger, so
that the net effect on the relic density is Boltzmann-suppressed. The bino-wino mixture does
not annihilate as efficiently as higgsinos, the corresponding relic density
turns to be larger than the Planck value, despite the co-annihilations.

\section{Benchmarks}
\label{sec:benchmarks}
To illustrate our results, we have selected five benchmark points with different
dark matter candidates and co-annihilation configurations. Each benchmark has a
relic density compatible with the Planck results within one standard
deviation. For each benchmark, the particle spectrum has been
computed at the one-loop level accuracy with {\sc Sarah}~4~\cite{Staub:2013tta}
and \textsc{SPheno} 3.3.8~\cite{Porod:2011nf}, while the
doubly-charged Higgs boson masses have been evaluated with the algorithm
described in Ref.~\cite{Basso:2015pka}. In practice, we  first chose $v_R$
to yield the $W_R$ boson heavy enough. We have then fixed the parameters relevant
for satisfying the Higgs sector constraints, namely $\lambda_{R}$, $v_S$ and
$\mathbf{h}_{RR}$ (to fix the properties of the doubly-charged Higgs bosons), as
well as $\tan\beta$ and $\tan\beta_{R}$ (for masses of the lightest
singly-charged, the $CP$-odd and the second $CP$-even Higgs bosons). The correct SM-like
Higgs-boson mass is retrieved by adjusting the stop masses and mixings.

For scenarios with a sneutrino LSP, we first set the sneutrino mass to a given
value before scanning over the slepton soft masses, $\lambda_{3}$ and $M_{2L}$.
This impacts the NLSP and the other co-annihilating particles as those
parameters respectively control the sneutrino, higgsino and $SU(2)_{\rm L}$-wino
masses. The scan is driven to obtain scenarios featuring a relic density
compatible with Planck data. In the co-annihilating higgsino case, we keep the
determination of $\lambda_{3}$ as the last step of the scan. The most important parameters
of the benchmark points are given in Table \ref{tab:parameters}. The full spectrum
information for all our benchmark scenarios can be obtained from
{\sc Inspire}~\cite{1712932}.

\begin{table}[!]
\begin{center}
\begin{tabular}{c c c c c c} 
\hline
\hline
Mass [GeV] & {\bf BP1} & {\bf BP2} & {\bf BP3} & {\bf BP4} & {\bf BP5} \\
\hline
$m_{W_R}$                &3509.8  &3509.8  &3369.5 &3696.9 &3369.5 \\
$m_{\tilde\chi^0_1}$     & 608.7  &696.8   &405.8  &429.0  &690.3 \\
$m_{\tilde\chi^0_2}$     & 707.8  &716.0   &417.4  &665.1  &718.6 \\
$m_{\tilde\chi^0_3}$     & 712.4  &716.1   &417.5  &665.3  &718.7 \\
$m_{\tilde\chi^0_4}$     & 712.4  &717.3   &419.1  &666.2  &719.8 \\
$m_{\tilde\chi^0_5}$     & 713.6  &851.4   &704.0  &677.5  &768.1 \\
$m_{\tilde\chi^{\pm}_1}$ & 699.3  &705.6   &411.7  &429.6  &712.5 \\
$m_{\tilde\chi^{\pm}_2}$ & 711.0  &715.3   &416.8  &663.3  &717.7 \\
$m_{\tilde\nu_1}$        & 278.1  &231.1   &387.5  &391.8  &1066.7 \\
$m_{\tilde\nu_2}$        & 662.2  &246.1   &1092.6 &658.8  & 1114.2\\
$m_{\tilde e_1}$         &814.6   &378.5   &1107.8 &470.4  &1150.7 \\
$m_{N_e}$         &136.8   &137.6   &131.9  &122.2  &131.8 \\
$m_{N_{\mu}}$     &158.4   &159.1   &152.7  &137.3  &152.6 \\
$m_{N_{\tau}}$    &719.6   &723.2   &694.9  &839.4  &707.6 \\
\hline
\hline
\end{tabular}
\caption{Relevant masses of the selected benchmark points. }
\label{tab:bp_mass}
\end{center}
\end{table}
\renewcommand{\arraystretch}{1.0}

\begin{table}
\begin{tabular}{c c c c c c}
\hline
\hline
& {\bf BP1} & {\bf BP2} & {\bf BP3} & {\bf BP4} & {\bf BP5} \\
\hline
$\tan \beta$ & $8$ & $7$ & $7$ & $7$ & $7$ \\
$\tan \beta_{R}$& $1.05$ & $1.045$ & $1.045$ & $1.04$ & $1.045$ \\
$V$ [TeV]& $7.5$ & $7.5$ & $7.2$ & $7.9$ & $7.2$ \\
$v_{S}$ [TeV]& $10$ & $7.2$ & $6.4$ & $7.8$ & $7.0$ \\
$\lambda_{3}$& $0.10$ & $0.14$ & $0.0915$ &  $0.12$ & $0.144$ \\
$\lambda_{R}$& $0.85$ & $0.90$ & $0.90$ & $0.90$ & $0.90$ \\
$M_{1}$ [GeV] & $400$ & $700$ & $550$ & $750$ & $700$ \\
$M_{2L}$ [GeV] & $900$ & $1000$ & $900$ & $412$ & $1200$ \\
$M_{2R}$ [GeV]& $900$ & $1000$ & $900$ & $1100$ & $650$ \\
\hline
\hline
\end{tabular}
\caption{Values of the most relevant parameters for the selected benchmark points. Here $V^{2}=v_{R}^{2}+v^{\prime 2}_{R}$. All benchmark points share the values  $\lambda_{4}=\lambda_{5}=0$, $\lambda_{L}=0.4$, $\lambda_{S}=-0.5$ and $\xi_{F}=-5\times 10^{5}$~GeV$^{3}$. The full sets of input parameters are available at Ref.~\cite{1712932}.}
\label{tab:parameters}
\end{table}

\begin{table}
\setlength\tabcolsep{2pt}
\begin{center}
\begin{tabular}{c c c c  c c}
\hline
\hline
  \multicolumn{6}{c}{Branching ratios in the {\bf BP1} scenario}\\
  \hline
  BR$({\tilde\chi^0_1}\to \nu{\tilde\nu_1})$         &1.00  &
    BR$({\tilde\chi^0_4}\to \nu{\tilde\nu_1})$       &1.00  &
    BR$({\tilde e_1}\to e{\tilde\chi^0_1})$          &0.96  \\
  BR$({\tilde\chi^0_2}\to \nu{\tilde\nu_1})$         &0.04  &
    BR$({\tilde\chi^0_5}\to \nu{\tilde\nu_1})$       &0.01  &
    BR$({\tilde e_1}\to e{\tilde\chi^0_2})$          &0.04  \\
  BR$({\tilde\chi^0_2}\to Z{\tilde\chi^0_1})$        &0.96  &
    BR$({\tilde\chi^0_5}\to Z{\tilde\chi^0_1})$      &0.99  &
    BR$(W_R\to \tilde\chi^0_i\tilde\chi^{\pm}_j)$    &0.23  \\
  BR$({\tilde\chi^0_3}\to \nu{\tilde\nu_1})$         &0.98  &
    BR$({\tilde\chi^{\pm}_1}\to W{\tilde\chi^0_1})$  &1.00  &
    BR$(W_R\to N\ell)$                               &0.17  \\
  BR$({\tilde\chi^0_3}\to Z{\tilde\chi^0_1})$        &0.02  &
    BR$({\tilde\chi^{\pm}_2}\to \tau{\tilde\nu_1})$  &1.00  &
    BR$(W_R\to qq^{\prime})$                         &0.50  \\
    \hline
    \hline
  \end{tabular}\vspace*{.2cm}

  \begin{tabular}{c c c c c c}
  \hline
  \hline
  \multicolumn{6}{c}{Branching ratios in the {\bf BP2} scenario}\\
  \hline
  BR$({\tilde\chi^0_1}\to \ell{\tilde \ell})$                         &1.00 &
    BR$({\tilde\chi^0_5}\to W{\tilde\chi^{\pm}_1})$                   &0.11 &
    BR$({\tilde\chi^{\pm}_2}\to \ell{\tilde\nu_1})$                   &0.06 \\
  BR$({\tilde\chi^0_2}\to \nu{\tilde\nu_1})$                          &1.00 &
    BR$({\tilde\chi^0_5}\to Z{\tilde\chi^0_4})$                       &0.02 &
    BR$({\tilde\chi^{\pm}_2}\to \tau{\tilde\nu_1})$                   &0.93 \\
  BR$({\tilde\chi^0_3}\to \nu{\tilde\nu_1})$                          &1.00 &
    BR$({\tilde\chi^0_5}\to h{\tilde\chi^0_1})$                       &0.03 &
    BR$({\tilde e_1}\to qq{\tilde\nu_{1,2}})$                         &1.00 \\
  BR$({\tilde\chi^0_4}\to \nu{\tilde\nu_1})$                          &0.69 &
    BR$({\tilde\chi^{\pm}_1}\to N{\tilde e_1})$                       &0.32 &
    BR$(W_R\to \tilde\chi^0_i\tilde\chi^{\pm}_j)$                     &0.22 \\
  BR$({\tilde\chi^0_4}\to \ell{\tilde\ell})$                          &0.30 &
    BR$({\tilde\chi^{\pm}_1}\to \tau{\tilde\nu_1})$                   &0.62 &
    BR$(W_R\to N\ell)$                                                &0.16 \\
  BR$({\tilde\chi^0_4}\to \ell\bar \ell{\tilde\chi^0_1})$             &0.01 &
    BR$({\tilde\chi^\pm_1}\to \ell\bar\nu^\prime_\ell\tilde\chi^0_1)$ &0.03 &
    BR$(W_R\to qq^{\prime})$                                          &0.50 \\
  BR$({\tilde\chi^0_5}\to \ell{\tilde\ell})$                          &0.83 &
    BR$({\tilde\chi^{\pm}_1}\to q\bar q^{\prime}{\tilde\chi^0_1})$    &0.03 &&\\
    \hline
    \hline
  \end{tabular}
  \caption{Relevant branching ratios for the {\bf BP1} and {\bf BP2} benchmark
  points.} \label{tab:bp_br}
\end{center}
\end{table}

\renewcommand{\arraystretch}{1.2}
\setlength\tabcolsep{10pt}
\begin{table}
  \begin{center}
  \begin{tabular}{c c c c c c}
  \hline
  \hline
  \multicolumn{6}{c}{Branching ratios in the {\bf BP3} scenario}\\
  \hline
  BR$({\tilde\chi^0_1}\to \nu{\tilde\nu_1})$                        &1.00&
    BR$({\tilde\chi^0_5}\to h{\tilde\chi^0_1})$                     &0.21&
    BR$({\tilde e_1}\to e{\tilde\chi^0_1})$                         &0.03\\
  BR$({\tilde\chi^0_2}\to \nu{\tilde\nu_1})$                        &1.00&
    BR$({\tilde\chi^{\pm}_1}\to \tau{\tilde\nu_1})$                 &0.85&
    BR$({\tilde e_1}\to e{\tilde\chi^0_5})$                         &0.97\\
  BR$({\tilde\chi^0_3}\to \nu{\tilde\nu_1})$                        &1.00&
    BR$(\tilde\chi^\pm_1\to\ell\bar\nu^\prime_\ell\tilde\chi^0_1)$  &0.05&
    BR$(W_R\to \tilde\chi^0_i\tilde\chi^{\pm}_j)$                   &0.22\\
  BR$({\tilde\chi^0_4}\to \nu{\tilde\nu_1})$                        &1.00&
    BR$({\tilde\chi^{\pm}_1}\to q\bar q^{\prime}{\tilde\chi^0_1})$  &0.10&
    BR$(W_R\to N\ell)$                                              &0.17\\
  BR$({\tilde\chi^0_5}\to Z{\tilde\chi^0_1})$                       &0.01&
    BR$({\tilde\chi^{\pm}_2}\to \tau{\tilde\nu_1})$                 &1.00&
    BR$(W_R\to qq^{\prime})$                                        &0.51\\
  BR$({\tilde\chi^0_5}\to Z{\tilde\chi^0_4})$                       &0.21&&&\\
  \hline
  \hline
  \end{tabular}\vspace*{.2cm}

  \begin{tabular}{c c c c c c}
  \hline
  \hline
  \multicolumn{6}{c}{Branching ratios in the {\bf BP4} scenario}\\
  \hline
  BR$({\tilde\chi^0_1}\to \nu{\tilde\nu_1})$                    &1.00 &
    BR$({\tilde\chi^0_4}\to Z{\tilde\chi^0_1})$                 &0.32 &
    BR$({\tilde\chi^{\pm}_2}\to Z\tilde\chi^{\pm}_1)$           &0.10\\
  BR$({\tilde\chi^0_2}\to Z{\tilde\chi^0_1})$                   &0.32 &
    BR$({\tilde\chi^0_5}\to W{\tilde\chi^{\pm}_1})$             &0.65 &
    BR$({\tilde\chi^{\pm}_2}\to q\bar q^{\prime}\tilde\chi^0_1)$&0.17\\
  BR$({\tilde\chi^0_2}\to W{\tilde\chi^{\pm}_1})$               &0.68 &
    BR$({\tilde\chi^0_5}\to Z{\tilde\chi^0_1})$                 &0.01 &
    BR$({\tilde\chi^{\pm}_2}\to q\bar q\tilde\chi^{\pm}_1)$     &0.10\\
  BR$({\tilde\chi^0_3}\to Z{\tilde\chi^0_1})$                   &0.01 &
    BR$({\tilde\chi^0_5}\to h{\tilde\chi^0_1})$                 &0.23 &
    BR$({\tilde\chi^{\pm}_2}\to \ell\bar\ell\tilde\chi^{\pm}_1)$&0.09\\
  BR$({\tilde\chi^0_3}\to h{\tilde\chi^0_1})$                   &0.32 &
    BR$({\tilde\chi^0_5}\to \ell\bar\ell{\tilde\chi^0_1})$      &0.02 &
    BR$({\tilde\chi^{\pm}_2}\to \nu\bar\nu\tilde\chi^{\pm}_1)$  &0.07\\
  BR$({\tilde\chi^0_3}\to W{\tilde\chi^{\pm}_1})$               &0.34 &
    BR$({\tilde\chi^0_5}\to q\bar q{\tilde\chi^0_1})$           &0.09 &
    BR$({\tilde e_1}\to e{\tilde\chi^0_1})$                     &1.00\\
  BR$({\tilde\chi^0_3}\to \ell\bar \ell{\tilde\chi^0_1})$       &0.02 &
    BR$({\tilde\chi^{\pm}_1}\to \ell{\tilde\nu_1})$             &1.00 &
    BR$(W_R\to \tilde\chi^0_i\tilde\chi^{\pm}_j)$               &0.22\\
  BR$({\tilde\chi^0_3}\to q\bar q{\tilde\chi^0_1})$             &0.28 &
    BR$({\tilde\chi^{\pm}_2}\to \ell{\tilde\nu_1})$             &0.18 &
    BR$(W_R\to N\ell)$                                          &0.16\\
  BR$({\tilde\chi^0_4}\to W{\tilde\chi^{\pm}_1})$               &0.67 &
    BR$({\tilde\chi^{\pm}_2}\to W\tilde\chi^0_1)$               &0.27 &
    BR$(W_R\to qq^{\prime})$                                    &0.50\\
    \hline
    \hline
  \end{tabular}\vspace*{.2cm}

  \begin{tabular}{c c c c c c}
  \hline
  \hline
  \multicolumn{6}{c}{Branching ratios in the {\bf BP5} scenario}\\
  \hline
  BR$({\tilde\chi^0_2}\to \ell\bar \ell{\tilde\chi^0_1})$           &0.11 &
    BR$({\tilde\chi^0_5}\to qq^{\prime}{\tilde\chi^{\pm}_1})$       &0.56 &
    BR$(\tilde\chi^\pm_2\to\ell\bar\nu^\prime_\ell\tilde\chi^0_1)$  &0.33  \\
  BR$({\tilde\chi^0_2}\to \nu\bar\nu{\tilde\chi^0_1})$              &0.21 &
    BR$(\tilde\chi^0_5\to\ell\nu^\prime_\ell\tilde\chi^{\pm}_1)$    &0.30 &
    BR$({\tilde e_1}\to e{\tilde\chi^0_1})$                         &0.32  \\
  BR$({\tilde\chi^0_2}\to q\bar q{\tilde\chi^0_1})$                 &0.68 &
    BR$({\tilde\chi^0_5}\to q\bar q{\tilde\chi^0_4})$               &0.10 &
    BR$({\tilde e_1}\to e{\tilde\chi^0_5})$                         &0.68  \\
  BR$({\tilde\chi^0_3}\to q\bar q{\tilde\chi^0_1})$                 &1.00 &
    BR$({\tilde\chi^0_5}\to \ell\bar\ell{\tilde\chi^0_4})$          &0.04 &
    BR$(W_R\to \tilde\chi^0_i\tilde\chi^{\pm}_j)$                   &0.22  \\
  BR$({\tilde\chi^0_4}\to \ell\bar \ell{\tilde\chi^0_1})$           &0.11 &
    BR$(\tilde\chi^\pm_1\to\ell\bar\nu^\prime_\ell\tilde\chi^0_1)$  &0.34 &
    BR$(W_R\to N\ell)$                                              &0.17  \\
  BR$({\tilde\chi^0_4}\to \nu\bar\nu{\tilde\chi^0_1})$              &0.21 &
    BR$({\tilde\chi^{\pm}_1}\to q\bar q^{\prime}{\tilde\chi^0_1})$  &0.66 &
    BR$(W_R\to qq^{\prime})$                                        &0.52  \\
  BR$({\tilde\chi^0_3}\to q\bar q{\tilde\chi^0_1})$                 &0.68 &
    BR$({\tilde\chi^{\pm}_2}\to q\bar q^{\prime}\tilde\chi^0_1)$    &0.67 &&\\
    \hline
    \hline
  \end{tabular}
  \caption{Relevant branching ratios for the {\bf BP3}, {\bf BP4} and {\bf BP5}
  benchmark points.} \label{tab:bp_br_bis}
\end{center}
\end{table}

Tables~\ref{tab:bp_mass}, \ref{tab:bp_br} and \ref{tab:bp_br_bis} show, for all
benchmarks, the
masses of the particles relevant for our study and their corresponding decay branching
ratios. In all cases, the $W_{R}$ boson decays mainly into a dijet system, a
lepton and a right-handed neutrino, or into a bidoublet higgsino-like
neutralino-chargino
pair, the decays into other electroweakino pairs being smaller or kinematically
forbidden. As right-handed neutrinos are Majorana fermions and doublets under
$SU(2)_{\rm R}$, $W_R$ boson decays into a lepton and a right-handed neutrino
are similar to its decays into a chargino and neutralino pair, up to phase-space
effects. The multiplicity then gives an extra factor of $4/3$ in favor of the
electroweakino channels. The typical branching fractions of the $W_R$ boson are
consequently
\begin{equation}
  {\rm BR}(W_R \to  j j) \sim 50\%\ , \qquad
  {\rm BR}(W_R \to  N \ell) \sim 16\%\ , \qquad
  {\rm BR}(W_R \to  \tilde\chi\tilde\chi) \sim 22\%\ ,
\end{equation}
with subdominant channels into various combinations of slepton pairs
($\sim 4\%$) or gauge and Higgs bosons ($\sim 4\%$). In all the setups except in
our second benchmark scenario {\bf BP2}, the lightest neutralino is either
stable or decays invisibly into a neutrino and a sneutrino LSP. $W_R$ decays
hence often lead to a significant amount of missing transverse energy, which
provide clear handles on LRSUSY.

\subsection{The BP1 scenario}
\label{subsec:bp1}
In this benchmark scenario, that is very close to the second scenario introduced
in Ref.~\cite{Frank:2017tsm}, the LSP is a right-handed tau sneutrino. It is
much lighter than any other superpartner, so that the relic density matches the
Planck value only by virtue of the sole sneutrino-pair annihilations. The
sneutrino mass lies in the 250--300~GeV mass window, in agreement with the
results of section~\ref{subsec:noco}, and the ${\bf h}_{RR}$ Yukawa texture
pushes the tau sneutrino to be the lightest one with a mass of 278~GeV. The
lightest neutralino is an admixture of the $\tilde{B}$ and $\tilde{W}_{R}^{0}$
gauginos and has a mass close to 600~GeV, while the next-to-lightest (charged
and neutral) electroweakinos are bidoublet-like higgsinos, with masses around
$700$~GeV.

In this scenario, the preferred $W_{R}$ boson supersymmetric decay
modes involve $\tilde{\chi}_{3,4}^{0}\tilde{\chi}^{\pm}_{1}$ and
$\tilde{\chi}_{2,5}^{0}\tilde{\chi}^{\pm}_{2}$ final states, each with a
branching ratio of about $5\%$.
The three $\tilde{\chi}_{1,3,4}^{0}$ neutralinos decay almost completely
invisibly to a $\nu\tilde{\nu}$ pair, although the $\tilde{\chi}^{0}_{3}$ state
subdominantly decays visibly into a $Z\tilde{\chi}^{0}_{1}$ pair (with a
branching ratio of $2\%$). This $Z\tilde{\chi}^{0}_{1}$ decay mode is
also the main  decay channel for the $\tilde{\chi}_{2,5}^{0}$ states. With branching ratios
greater than 95\%, this can further give rise to the production of opposite-sign
same-flavor lepton pairs at colliders. The two lighter charginos respectively
decay into $W^{\pm}\tilde{\chi}^{0}_{1}$ and $ \tau^{\pm}\tilde{\nu}$ systems,
the lepton flavor in the latter case being connected to the LSP nature.
Altogether such a spectrum has a good chance to copiously produce hard leptons
at colliders, in association with missing energy, which could provide handles on
the model.

\subsection{The BP2 scenario}
\label{subsec:bp2}
Our second benchmark features two light right-handed sneutrinos, of the tau and
electron flavors, so that the reproduction of the right relic density value
largely relies on co-annihilations. As mentioned in section~\ref{%
subsec:sneutrinos}, the LSP mass is lower than in setups like the one of our
benchmark {\bf BP1}, where co-annihilations are negligible. The lighter
electroweakinos, being of a bidoublet higgsino nature, are heavier with a mass
close to $700$~GeV, whilst the lightest charged slepton is a right-handed
selectron with a mass of $378$~GeV. In addition, the soft mass configuration
yielding the sneutrino hierarchy additionally makes the stau states heavier.

Here, the supersymmetric $W_{R}$ boson decays mainly involve
$\tilde{\chi}_{1,4}^{0}\tilde{\chi}^{\pm}_{2}$ and
$\tilde{\chi}_{2,3}^0\tilde{\chi}^{\pm}_{1}$ final states, each with a branching
fraction close to $5\%$. The lightest neutralino always decays into an
$e^{\pm}\tilde{e}_{R}^{\mp}$ system, which is also a relatively dominant decay mode of
the $\tilde{\chi}_{4}^{0}$ state with a branching ratio of 30\% (muonic
contributions being subdominant). In contrast, the $\tilde{\chi}_{2,3}^{0}$
neutralinos decay invisibly. Because of the structure of the Yukawa
couplings, the higgsino-like charginos decay mostly into $\tau\tilde{\nu}$
systems, although $\tilde{\chi}^{\pm}_{1}$ decay into a
$N_{e}\tilde{e}^{\pm}_{R}$ final state is significant too. Finally, the
selectrons often appearing at the end of the decay chain dominantly decay via a
virtual $W_{R}$ boson into a $jj\tilde{\nu}_{e}$ system.

This benchmark point can be probed at colliders through a signature involving
multilepton final states, the corresponding rate being large enough. The signal
often contains electrons, by the nature of the spectrum featuring light electron
sneutrino and selectron. Tau leptons are also largely produced, in particular
when decays involve charginos ($\tilde{\chi}^{\pm}_{1/2}$).

\subsection{The BP3 scenario}
\label{subsec:bp3}
For our third benchmark, we picked a scenario where the LSP is a right-handed
tau sneutrino almost degenerate with a set of bidoublet higgsinos. Contrary to
the {\bf BP2} scenario, the LSP is moderately heavier with a mass of 387~GeV,
and the right relic density is once again obtained thanks to co-annihilations.
The higgsinos are about $20$--$35$~GeV heavier.

The supersymmetric decays of the $W_{R}$ boson are similar to the previous
cases. The main electroweakino channels, with a branching ratio of $5.5\%$ each,
involve $\tilde{\chi}^{0}_{1,4}\tilde{\chi}^{\pm}_{2}$ and
$\tilde{\chi}^{0}_{2,3}\tilde{\chi}^{\pm}_{1}$ systems. The four lighter
neutralinos all decay invisibly into a sneutrino/neutrino pair, and the two
lighter charginos decay mostly into $\tau\tilde{\nu}$ systems by virtue of the
large tau Yukawa coupling, the other channels being three-body and involving
virtual $W$ bosons. Although the decays of heavier neutralinos are visible, they
are barely produced via intermediate $SU(2)_{\rm R}$ gauge bosons. This scenario
hence manifests itself at colliders through an enhanced production of tau
leptons in association with missing transverse energy.

\subsection{The BP4 scenario}
\label{subsec:bp34}
Our fourth scenario has very different features from the previous one, although
both have a sneutrino LSP giving rise to the right relic abundance through
co-annihilations. The latter, however, involves this time $SU(2)_{\rm L}$ wino-like
neutralinos and charginos, and the LSP is here an electron sneutrino. This
means that the associated collider signature involves electrons instead of taus.
The LSP sneutrino mass is of 391~GeV, and the lighter neutralinos and charginos
have masses of about 430~GeV. This mass difference, slightly larger than for the
{\bf BP3} scenario, is necessary for yielding a relic density matching the
Planck results, because of the existence of additional DM annihilation
subprocesses into right-handed neutrinos via $t$-channel neutralino exchanges.
As will be shown below, this annihilation channel is crucial for DM indirect
detection. Finally, the bidoublet higgsinos are heavier, with masses lying
around $740$~GeV.

The neutral and charged light winos decay into $\nu\tilde{\nu}$ and
$e\tilde{\nu}$ pairs, respectively, as those are the only possible decay modes.
The heavier neutralinos and charginos feature in addition significant branching
ratios into the $W\tilde{\chi}_{1}$ and $Z\tilde{\chi}_{1}$ modes, so that
$W_R$ boson production and supersymmetric decay into electroweakinos (with
branching fractions similar to the other benchmarks) could lead to an important
production of hard jets and leptons, and electrons in particular, at colliders.
Representative signatures of this benchmark feature the intermediate presence
of weak bosons whose reconstruction could provide interesting handles to unravel
the signals.

\subsection{The BP5 scenario}
\label{subsec:bp5}
In contrast to all other scenarios, the lighter superpartners are all bidoublet
higgsinos and the numerous existing co-annihilation modes allow for a viable
neutralino DM candidate with a mass of about $700$~GeV. This rather heavy
spectrum consists in a perfect example of stealth supersymmetry. As any new
gauge boson or colored superpartner is heavy enough for their production rate to
be suppressed, any potential collider signal becomes hard to get. Even when
considering cascades such as those originating from the production of a single
$W_R$ boson, a large integrated luminosity would be necessary to observe any
signal. The resulting rate is reasonable enough to give hope for
detection.

The lightest neutralinos and charginos being nearly degenerate, their decay
proceed via three-body channels. The $\tilde{\chi}^{0}_{2,4}$ states hence give
rise to $\nu\overline{\nu}\tilde{\chi}_{1}^{0}$ (21\%), $\ell^{+}\ell^{-}
\tilde{\chi}_{1}^{0}$ (11\%) and $jj\tilde{\chi}_{1}^{0}$ (68\%) final states
through an off-shell $Z$ boson, whilst the $\tilde{\chi}^{0}_{3}$ neutralino
decays in contrast dominantly through an off-shell $h$ boson into a $b\bar b
\tilde{\chi}^{0}_{1}$ final state. Similarly, chargino decays involve a virtual
charged $W$ boson instead. The main $W_R$ boson signature hence consists in a
production of numerous leptons, jets and  missing energy.

\section{Implications for dark matter indirect detection}
\label{sec:indirect}
In this section we discuss the implications of indirect DM searches on the
representative LRSUSY scenarios reproducing the Planck results introduced in the
previous section. Since the center of the Milky Way and dwarf spheroidal galaxies (dSPhs) are enriched in dark matter, various indirect searches for DM focus on these regions
of the universe to extract a DM signal using various classes of cosmic rays, see
{\it e.g.} Refs.~\cite{Bringmann:2012ez,2012CRPhy..13..740L,2010arXiv1001.4086P}.
Calculations associated with neutral particles ({\it i.e.}  photons and neutrinos) do
not suffer from propagation uncertainties, so that gamma rays or neutrino fluxes
can be efficiently used to probe DM annihilation. In particular, gamma-ray flux
measurements are widely considered to constrain DM annihilation into varied SM
states, thanks to
the ease of their detection~\cite{Bringmann:2012ez}. Constraints on specific
final states have been recently evaluated by the Fermi collaboration, using both
the continuum gamma-ray fluxes originating from dSPhs~\cite{Ackermann:2015zua}
or from the galactic center~\cite{TheFermi-LAT:2017vmf}. In addition, the
implications on DM annihilation into new physics final states like right-handed
neutrinos have also been studied~\cite{Campos:2017odj}.

Constraints on the late-time thermally-averaged DM annihilation cross section
($\langle\sigma v \rangle$) put forward by the Fermi experiment all assume that
DM annihilates into a particular SM channel. However, several annihilation
channels are generally open, depending on the model itself and on the exact
value of its free parameters. Furthermore, new physics annihilation modes can be
open too, like in LRSUSY scenarios where DM
often annihilates into right-handed neutrinos. For illustrative purposes, we
start by estimating the prompt gamma-ray flux originating from DM annihilation
in the context of our characteristic benchmark scenario {\bf BP4}, where DM
annihilation into right-neutrinos is possible along with multiple other
channels. We next attempt to put an upper bound on the DM annihilation
cross section from the observation by Fermi-LAT of the gamma-ray spectrum issued
from 15 dSPhs, in the 0.5-500 GeV energy range~\cite{Ackermann:2015zua}. We
ignore any potential constraints that could emerge from measurements of the
gamma-ray flux originating from the galactic center, as the expectation is
comparable but plagued by a large background stemming from other astrophysical
processes~\cite{TheFermi-LAT:2017vmf}.

The observed gamma-ray flux ($\Phi_{\gamma}$) is related to the DM annihilation
cross section as
\begin{equation}
  \frac{{\rm d} \Phi_{\gamma}}{{\rm d} E} (E_{\gamma}, \Delta \Omega) = 
    \frac{1}{4 \pi}\frac{\langle \sigma v \rangle}{2 m_{\rm DM}^2}
    \frac{{\rm d}N_{\gamma}}{{\rm d}E_{\gamma}}\int {\rm d}\Omega
    \int_{\rm l.o.s} {\rm d}r \rho^2(r) \ ,
\end{equation}
where the differential gamma-ray flux embedded in a solid angle $\Delta \Omega$
has been equated to the prompt gamma-ray flux generated from the annihilation of
a pair of DM particles of mass $m_{\rm DM}$. The prefactor of the integral
includes the thermally-averaged DM annihilation cross section $\langle \sigma v
\rangle$ today and in the relevant galaxies, and the differential number density
of photons within an energy bin of size ${\rm d}E$, ${\rm d}N_{\gamma} / 
{\rm d}E_{\gamma}$. The two integrals in the right-hand side represent the
so-called astrophysical $J$-factor and account for the squared density of dark
matter $\rho$ along the line of sight (l.o.s). We have relied
on NFW profiles to estimate the $J$ factor~\cite{Ackermann:2015zua,
Workgroup:2017lvb}, which corresponds to an uncertainty of at most 30\%~\cite{
Ackermann:2015zua}. Accounting for a different density profile is however not
expected to significantly affect the results~\cite{Ackermann:2015zua}.
The estimation of the contribution from the inverse Compton effect 
is plagued by a different uncertainty, most notably the uncertainty in the diffusion 
coefficient, which can affect the final photon spectrum up to one order of magnitude \cite{McDaniel:2017ppt}.
As we solely consider prompt
gamma-ray production at the source, any other contribution such as those
stemming from inverse Compton scattering, bremsstrahlung and synchrotron
radiation are neglected. While the latter two are subdominant, inverse Compton
scattering at dwarf galaxies is potentially relevant but necessitates a
detailed modelling of electron and positron propagation to the observer. It will
therefore be omitted from our calculations.

\begin{figure}
  \begin{center}
    \includegraphics[width=.48\columnwidth]{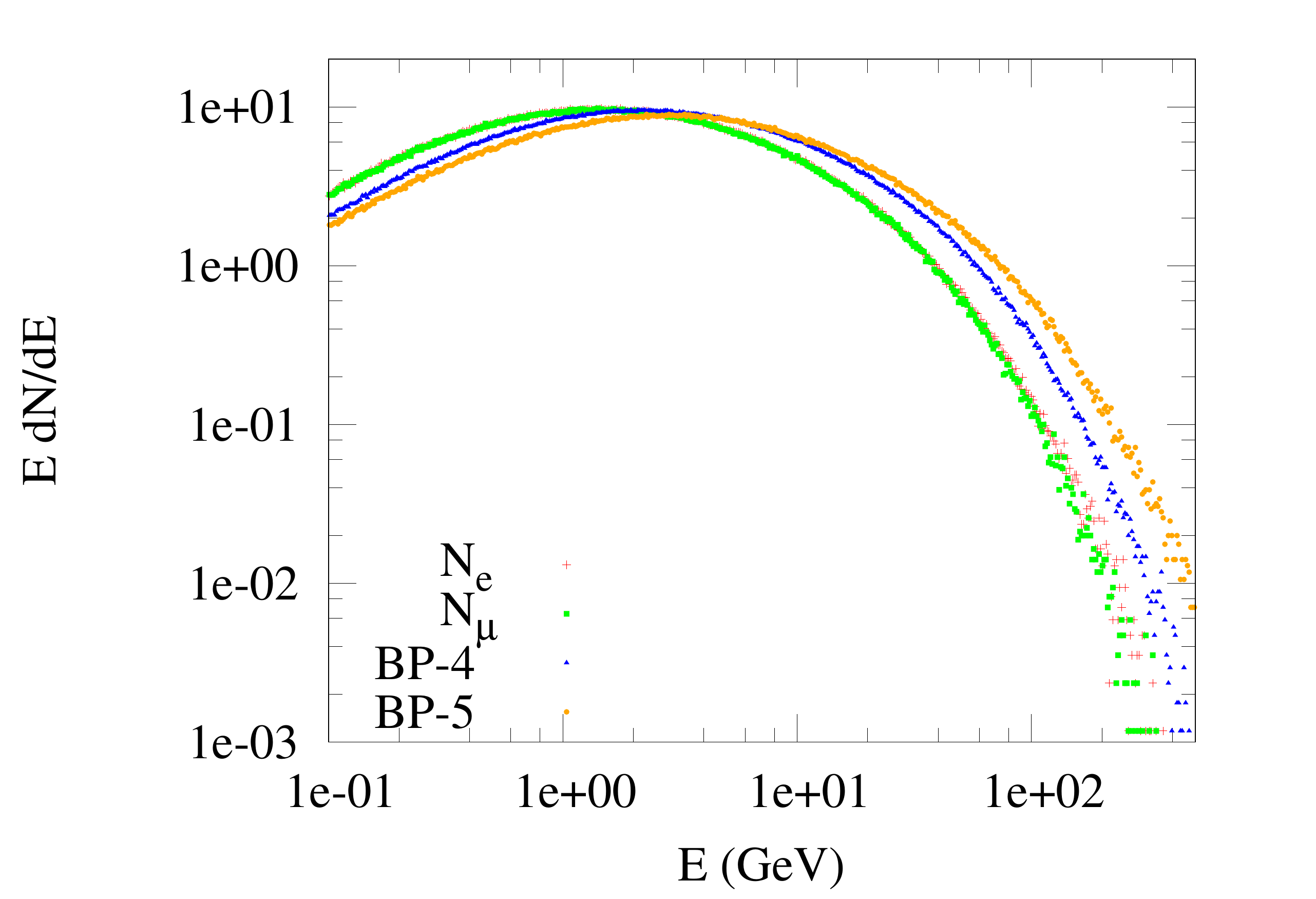}
    \includegraphics[width=.48\columnwidth]{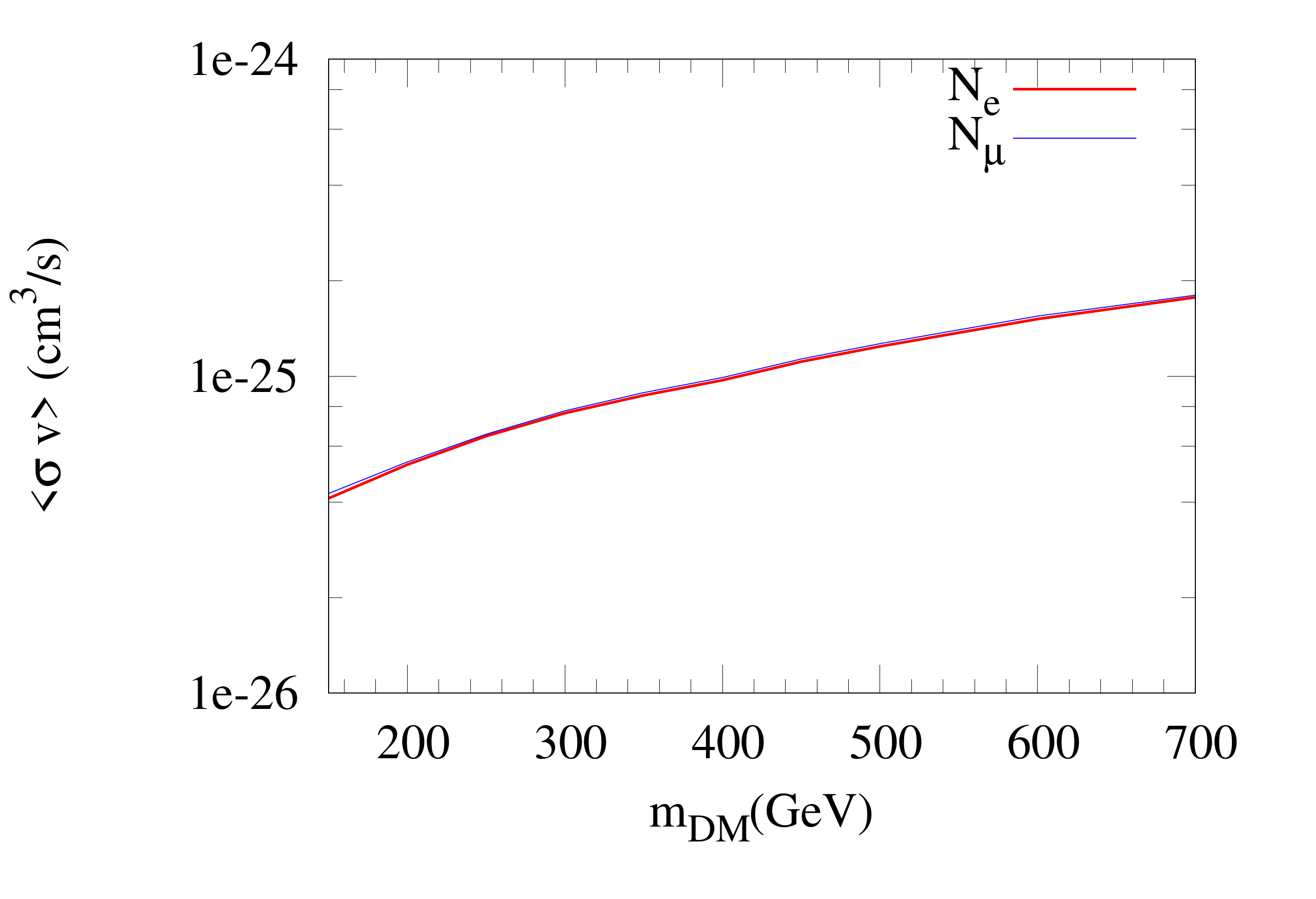}
    \caption{Photon number density spectrum (left) as derived from the primary
      gamma-ray flux originating from DM annihilation into right-handed
      neutrinos, shown as a function of energy. Results are presented for a
      showcase scenario in which $m_{\rm DM} = 400$~GeV and for electron (red) and
      muon (green) neutrinos of 130~GeV, as well as for both the {\bf BP4}
      (blue) and {\bf BP5} (orange) scenarios. The upper limits on the DM
      annihilation cross-section
      at the $95\%$ confidence level, derived from Fermi dSPhs data, are also
      shown for both the right-handed electron and muon channels as a function
      of the DM mass (right).}
        \label{fig:dm}
  \end{center}
\end{figure}

In the left panel of Figure~\ref{fig:dm}, we present the (normalized) prompt
gamma-ray flux produced by the annihilation of pairs of DM particles into
electronic and muonic right-handed neutrinos $N_e$ (red) and $N_{\mu}$ (green),
for a DM mass of $m_{\rm DM} = 400$~GeV and neutrino masses of 130~GeV. The
particle shower generated from the neutrino decays,
\begin{equation}
  N_e \rightarrow e^{\pm} q q' \qquad\text{and}\qquad
  N_\mu \rightarrow \mu^{\pm} q q' \ ,
\end{equation}
has been simulated with \textsc{Pythia 8}~\cite{Sjostrand:2014zea}. In addition,
we have also shown results for the {\bf BP4} (blue) and {\bf BP5} (orange)
representative benchmark scenarios featuring masses in the same ball park. In
the right panel of the figure, we derive 95\% confidence level limits on the
dark matter annihilation cross section in neutrinos from the  Fermi results,
making use of the publicly available likelihood calculator
\textsc{GamLike}~\cite{Workgroup:2017lvb}. We then apply these findings to our
{\bf BP4} scenario. From all the competing subprocesses, the
driving component turns out to be the annihilation of sneutrinos in a pair of
$W$ bosons ($\sim 45\%$), and $Z$ bosons ($\sim 21\%$), as well as into Higgs
bosons ($\sim 15\%$), followed by right-handed neutrinos ($\sim 10\%$) and top quarks ($ \sim 8\%$). 
The corresponding annihilation cross section at present time being $4.3 \times 10^{-27}~{\rm
cm^3/s}$, the scenario turns out to be safe relative to  Fermi observations
that exclude, at the 95\% confidence level, cross sections of $9.0 \times
10^{-26}~{\rm cm^3/s}$.

We obtain a similar conclusion for our other benchmark scenarios with a similar
dark matter mass. In the {\bf BP5} scenario with a slightly heavier DM candidate
of 690~GeV, we get a slightly harder gamma-ray spectrum (left panel of
Figure~\ref{fig:dm}), the dominant process driving the results being the
annihilation of pairs of neutralinos into charged ($\sim 45\%$) and neutral
($\sim 33\%$) electroweak bosons, as well as into $hZ$ associated pairs
($\sim 12 \%$). The corresponding thermally-averaged annihilation cross section
at the present epoch reads $1.60 \times 10^{-28}~{\rm cm^3/s}$, which is again
well below the computed  Fermi exclusion of $1.7 \times 10^{-25} {\rm cm^3/s}$.

\section{Resonant production of charginos and neutralinos at colliders}
\label{sec:collider}
\subsection{Generalities and analysis definition}
\label{subsec:general_lhc}
A robust experimental confirmation of the existence of a left-right symmetry
would incontrovertibly consist in the discovery of a charged $SU(2)_{\rm R}$
gauge boson $W_R$. The ATLAS and CMS collaborations have therefore extensively
searched for varied $W_R$ boson signals,
in particular in the $\ell\ell jj$ channel. $W_R$ boson masses ranging up to
about 4.5~TeV have been excluded, assuming that at least one of the right-handed
neutrinos $N$ is lighter than the new boson~\cite{Sirunyan:2018pom}, the exact
value of the bound being subjected to the $N$ mass. More precisely, there is
no constraint when $m_{N}\simeq m_{W_R}$ or when $m_{N}$ is below a certain
threshold, ({\it e.g.} the $m_{W_R} > 3$~TeV region is unconstrained if
$m_{N}\lesssim 150$~GeV)~\cite{Sirunyan:2018pom}. In this case, much more
robust bounds, $m_{W_R}\gtrsim$ 3.5 TeV, are obtained from the analysis of dijet
probes for extra gauge bosons~\cite{Aaboud:2017yvp,Sirunyan:2018xlo}. All these
bounds can, however, be relaxed in an LRSUSY context, thanks to the presence of
additional (supersymmetric) decay modes of the $W_R$ boson.

In the scenarios investigated in this work, we accommodate a correct DM relic
density by enforcing the presence of multiple neutralino and chargino states
that are slightly heavier than the LSP. As described in
section~\ref{sec:benchmarks}, this yields to new possible decay channels for the
$W_R$ boson, associated with a combined branching ratio that could be as large
as 25\%. This opens up the opportunity to look for a $W_R$ boson through typical
electroweakino searches targeting multileptonic final states exhibiting a large
amount of missing energy $\slashed{E}_{T}$. As such signals are absent in
minimal non-supersymmetric left-right extensions of the SM, they consist,
together with the would-be observation of a $W_R$ boson, of clear evidence for
LRSUSY. The existence of these new $W_R$ boson signals is also responsible for
the reduction of the reach of the classical $W_R$ boson searches, as the
branching ratios relevant for the latter are automatically reduced by virtue of
the new decay modes. Moreover, these new channels, even if featuring missing
energy, also provide the opportunity to reconstruct a $W_R$ boson mass from the
kinematic thresholds featured by numerous transverse variables.

To illustrate this point, we have followed the CMS multilepton analysis
dedicated to MSSM electroweakino searches at ${\sqrt s}=13$ TeV  of Ref.~\cite{Sirunyan:2017lae}. 
This
experimental study includes various signal regions defined according to the
final-state lepton (electron and muon) multiplicity $N_\ell$ and tau lepton
multiplicity $N_\tau$, the number of opposite-sign same-flavor pairs of leptons
$N_{\rm OSSF}$, same-flavor pairs of leptons $N_{\rm SF}$ and opposite-sign
pairs of leptons $N_{\rm OS}$, as well as the value of several kinematic
variables
like the missing transverse energy $\slashed{E}_{T}$, lepton or dilepton
transverse momenta, the transverse mass $M_T$ of systems made of a lepton and
the missing momentum, the stransverse mass $M_{T2}$ and the invariant mass of
various dilepton systems $M_{\ell\ell}$. 

The reason for which we choose to recast the CMS multilepton analysis of Ref.~\cite{Sirunyan:2017lae}
stems from the nature of the final state originating from the decay of a chargino-neutralino system induced by a $W_R$-boson exchange.
However, the electroweakinos originating from resonance production of a much heavier
particle are expected to be more boosted, resulting in harder kinematic distributions of 
the corresponding MSSM signal. 
In order to maximize the signal versus background ratio, we therefore narrow down
our investigations to the signal regions in which the ${\slashed{E}_T}$, $M_T$,
$M_{\ell\ell}$ and $M_{T2}$ kinematical variables are imposed to be larger than
some threshold values, instead of lying within a specific window. The overflow bins that we focus on
are not necessarily the most sensitive signal regions for MSSM electroweakino searches, but prove to be 
so in our case.
Owing to the Majorana nature of the
neutralinos, signal events featuring  same-sign dileptons are expected to be
copiously produced. However, it turns out that the sensitivity of their
corresponding parameter regions is smaller than the one of regions requiring three or four
leptons. The properties of the signal regions that are most suitable for probing
the different types of spectra considered in
this work are summarized in Table~\ref{tab:def_sr}.
Whilst it is clear that combining the strengths of several regions
would increase the overall sensitivity, this task requires to obtain non-public
information on uncertainty correlations, so that we prefer to be conservative
and focus on one region at a time.

\begin{table}
\begin{center}
  \begin{tabular}{c c}
  \hline
  \hline
  SR & Requirements \\
  \hline
  {\bf A44}  & $N_\ell=3$, $N_\tau=0$, $N_{\rm OSSF}\ge 1$, $M_T > 160$~GeV,
    $\slashed{E}_T\ge200$~GeV, $M_{\ell\ell}\ge 105$~GeV\\
  {\bf C18}  & $N_\ell=2$, $N_\tau=1$, $N_{\rm OSSF}=1$, $M_{T2}>100$~GeV,
    $\slashed{E}_T\ge200$~GeV, $M_{\ell\ell}\ge 105$~GeV\\
  {\bf D16}  & $N_\ell=2$, $N_\tau=1$, $N_{\rm OS}=1$, $N_{\rm SF}=0$,
    $M_{T2} > 100$~GeV, $\slashed{E}_T \ge 200$~GeV\\
  {\bf G05}  & $N_\ell\ge 4$, $N_\tau=0$, $N_{\rm OSSF}\ge 2$,
    $\slashed{E}_{T} \ge 200$~GeV\\
  {\bf H04}  & $N_\ell\ge 4$, $N_\tau=0$, $N_{\rm OSSF} < 2$,
    $\slashed{E}_{T} \ge 150$~GeV\\
    \hline
    \hline
  \end{tabular}
  \caption{Definition of the signal regions (SR) of the CMS analysis of
    Ref.~\cite{Sirunyan:2017lae} that we use as potentially best probes of our
    cosmologically-favored LRSUSY scenarios.}
\label{tab:def_sr}
\end{center}
\end{table}

More in details, all the signal regions under category-{\bf A} require the presence of three
light-flavored charged leptons ($\ell\equiv e$, $\mu$), with transverse momenta
greater than 25 (20)~GeV and 15 (10)~GeV for the leading and subleading leptons
respectively, in the electron (muon) case. The pseudorapidity of each electron
(muon) is moreover imposed to satisfy $|\eta_{\ell}| < 2.5$ (2.4). The three
selected lepton candidates must form at least one opposite-sign same-flavor
(OSSF) pair of leptons, and feature a trilepton invariant mass satisfying
$|m_{3\ell}-m_Z| > 15$~GeV. In addition, for the {\bf A44} signal region
the invariant mass of the dilepton system
constructed from the OSSF lepton pair that is as compatible as possible with a
$Z$ boson decay is enforced to be larger than 105~GeV, and the transverse mass
of the system made of the missing momentum and the third lepton is constrained
to be larger than 160~GeV. Finally, one demands that the missing energy
$\slashed{E}_{T} > 200$~GeV.

Signal regions under category-{\bf C} are dedicated to final-states that feature two
light-flavored charged leptons which fulfill similar requirements as for the ones in category-{\bf A}, 
and one hadronic tau with a transverse momentum $p_T > 20$~GeV and
a pseudorapidity satisfying $|\eta_\tau| < 2.3$. For signal region {\bf C18}, the two light-flavored leptons
are enforced to form an OSSF lepton pair with an invariant mass $M_{\ell\ell}$
larger than 105~GeV. The missing transverse energy and the event stransverse
mass $M_{T2}$ are finally imposed to be respectively larger than 200 and
100~GeV.

Signal regions under category-{\bf D} also focus on a topology featuring two
light-flavored charged leptons and one hadronic tau. In addition to the previous
requirements on the leptons, the $p_T$ threshold on the leading light-flavored
lepton is increased to 25~GeV when it is a muon and when the subleading
light-flavored lepton is an electron. These two leptons are moreover required
to be of opposite signs and of different flavors. Event selection for {\bf D16} require at
least 200~GeV of missing transverse energy and a stransverse mass larger than
100~GeV, the variable being constructed from the $e\mu$ pair.

Signal regions under category-{\bf G} focus on final states containing at least four
light-flavored charged leptons (with transverse momentum and pseudorapidity
requirements as above), and no hadronic tau. One requires these leptons to form
at least two OSSF lepton pairs and for {\bf G05}, the events should satisfy $\slashed{E}_T>200$~GeV.
Signal regions under category-{\bf H} are similar, except that they require 
at most one OSSF lepton pair and for {\bf H04}, $\slashed{E}_T>200$~GeV.

In the following, we estimate the LHC sensitivity to our LRSUSY setups by
investigating how the five above-mentioned signal regions are populated by the
$W_R$ boson mediated electroweakino signal. To this aim, we have used a {\sc Sarah {\color{magenta}4}}
implementation of the model~\cite{Staub:2013tta,Basso:2015pka} to be able both
to generate the particle spectrum with \textsc{SPheno 3}~\cite{Porod:2011nf}, as
already above-mentioned, and
to export the model information under the form of an
LRSUSY UFO library~\cite{Degrande:2011ua} to be used with
{\sc MadDM {\color{magenta}2.0}}~\cite{Backovic:2015cra} (for DM calculations) and
{\sc MG5\_aMC@NLO} v2.5.5~\cite{Alwall:2014hca} (for hard scattering LHC event
generation). Our computations rely on the leading order set of NNPDF 2.3 parton
distribution functions~\cite{Ball:2014uwa}.
The simulation of the LHC QCD environment has been achieved with
\textsc{Pythia 8}~\cite{Sjostrand:2014zea} and we have made use of {\sc
MadAnalysis} 5~\cite{Conte:2012fm,Conte:2014zja} to automatically handle the
impact of the CMS detector with {\sc Delphes}~3~\cite{deFavereau:2013fsa} and
event reconstruction with {\sc FastJet} 3.3.0~\cite{Cacciari:2011ma}. We have
reinterpreted the results of the CMS analysis of Ref.~\cite{Sirunyan:2017lae}
by relying on the {\sc MadAnalysis}~5 reimplementation of this
analysis~\cite{cms:multl}, available from the \textsc{MadAnalysis} Public
Analysis Database~\cite{Dumont:2014tja,Conte:2018vmg}.

\subsection{Results}

\begin{table}
\begin{center}
  \begin{tabular}{cccccccc}
  \hline
  \hline
   & {\bf BP1} & {\bf BP2} & {\bf BP3} & {\bf BP4}& {\bf BP5} & SM & Obs.\\
   \hline
   $\sigma(pp\to W_R)$ (fb) &38.12   &38.12  &51.54  &25.58 &51.54   & -- & --\\ 
   \hline
   {\bf A44} &0.75  &0.90  &0.93   &2.07   &0.42  &$2.5\pm 0.8$ & 0\\
   {\bf C18} &0.78  &0.28  &1.30   &0.27   &0.24  &$1.9\pm 0.7$  & 1\\
   {\bf D16} &0.85(0.57)  &0.94(0.61)  &0.77(0.54)   &0.43(0.48)   &0.43(0.46)  &$0.06\pm 0.05$ & 0\\
   {\bf G05} &0.02  &0.11  &0.09   &1.38   &0.09  &$0.97\pm 0.32$  &  0\\
   {\bf H04} &0.03  &0.66  &0.24   &1.81   &0.20  &$1.9\pm 0.6$ &  1\\
   \hline
   \hline
  \end{tabular}
  \caption{Number of signal events populating the various
    signal regions of interest, normalized to an integrated luminosity of
    35.9~${\rm fb^{-1}}$ at 13 TeV LHC
    collisions. We show the $W_R$ boson production rate (first line) and
    the background, along with the number of observed events, as
    reported by the CMS collaboration in Ref.~\cite{Sirunyan:2017lae}. The numbers in parentheses
    following the event counts for the most sensitive signal region ({\bf D16}) indicate the
    CLs values computed with {\sc MadAnalysis}~5~\cite{Conte:2012fm,Conte:2014zja}.}
  \label{tab:nev_sig}

  \vspace*{.3cm}

  \begin{tabular}{c ccccc}
  \hline
  \hline
    &  {\bf BP1} & {\bf BP2} & {\bf BP3} & {\bf BP4} & {\bf BP5} \\
    \hline
    ${\cal L}$ & $300 (3000)~{\rm fb}^{-1}$& $300 (3000)~{\rm fb}^{-1}$&
      $300 (3000)~{\rm fb}^{-1}$& $300 (3000)~{\rm fb}^{-1}$&
     $300 (3000)~{\rm fb}^{-1}$\\
   \hline
   {\bf A44} &0.74 (0.91) &0.88 (1.09) &0.91 (1.13) &1.90 (2.50) &0.42 (0.51)\\
   {\bf C18} &0.87 (1.08) &0.32 (0.39) &1.39 (1.79) &0.31 (0.38) &0.28 (0.33)\\
   {\bf D16} &2.55 (7.35) &2.69 (7.82) &2.41 (6.91) &1.74 (4.70) &1.74 (4.70)\\
   {\bf G05} &0.04 (0.06) &0.23 (0.32) &0.19 (0.27) &2.22 (3.82) &0.19 (0.27)\\
   {\bf H04} &0.04 (0.05) &0.81 (1.06) &0.31 (0.39) &2.02 (2.85) &0.26 (0.32)\\
   \hline
   \hline
  \end{tabular}
  \caption{Statistical significance ${\mathcal S} = S/\sqrt{S+B+\sigma_B^2}$
    when 300 and 3000~fb$^{-1}$ of 13 TeV LHC collisions are considered, for the
    the five benchmarks under investigation. The level of background uncertainty
    $\sigma_B/B$ is assumed equal to the 35.9~fb$^{-1}$ case of
    Ref.~\cite{Sirunyan:2017lae}.}
  \label{tab:nev_sig1}
\end{center}
\end{table}

 Using the methods presented at
the end of section~\ref{subsec:general_lhc}, we present our results, for  35.9~fb$^{-1}$ of
LHC collisions at a center-of-mass energy of 13~TeV, in Tables~\ref{tab:nev_sig}
and \ref{tab:nev_sig1}. Table~\ref{tab:nev_sig} shows our predictions for the
number of signal events expected to populate the five considered signal
regions, for an integrated luminosity of 35.9~${\rm fb^{-1}}$ at 13 TeV LHC
collisions. Those results can be compared with the SM expectation as extracted
from Ref.~\cite{Sirunyan:2017lae} (last column). Table~\ref{tab:nev_sig1} shows
estimates for the signal significance defined as
\begin{equation}
  {\mathcal S} = \frac{S}{\sqrt{S+B+\sigma_B^2}} \ ,
\end{equation}
where $S$ and $B$ stands for the number of signal and background events
populating a given signal region, and $\sigma_B$ stands for the background
uncertainty. Our results are provided for two different luminosity
configurations of the LHC, namely 300 and 3000~fb$^{-1}$, and for the five
different benchmark points and the five considered signal regions. We have
estimated the background contributions by rescaling the expectation of
Ref.~\cite{Sirunyan:2017lae}, and we have assumed the same uncertainty as for
$35.9~{\rm fb}^{-1}$.

From the branching ratio
information enlisted in Tables~\ref{tab:bp_br} and \ref{tab:bp_br_bis}, it is
clear that multilepton production from $W_R$ boson decays is important.
Trilepton final states (probed by the {\bf A44}, {\bf C18} and {\bf D16}
regions) can, in principle, originate from both $W_R\to\widetilde\chi^0_i\widetilde\chi^\pm_j$
and $W_R\to N\ell$ decays. However, in our present scenario, the left-right mixing 
being extremely small, all the right-handed
neutrinos entirely decay via off-shell $W_R$ 
into the three body decay mode $\ell qq^{\prime}$. Thus three or more leptons in the final state
can only arise substantially from $W_R$ decay into the neutralino-chargino pairs 
and the subsequent cascades.
In the context of the {\bf BP1} scenario, all three trilepton regions are
roughly similarly populated, with a slight preference for the {\bf D16} region.
Leptons mostly originate from the decays of the weak bosons produced in the
electroweakino cascade decays, although they can directly stem, in some rarer
cases, from a $\tilde\chi^{\pm}_2$ decay into a sneutrino LSP along
with an extra (often tau) leptons. Trilepton production in the {\bf BP2}
context is slightly enhanced compared with the {\bf BP1} scenario, thanks to the
presence of a light selectron in the spectrum allowing higgsinos to produce more
electrons in their decays. Tau production is additionally ensured via chargino
decays, so that the {\bf D16} and {\bf A44} regions turn out to be better to
probe the signal. The {\bf BP3} scenario features a more compressed
spectrum and understandably, one requires a lighter $W_R$ in order to obtain 
comparable event rates. $\tau$-enriched final states are more prominent 
in this case because of the large decay branching ratios of both $\widetilde\chi_1^{\pm}$ 
and $\widetilde\chi_2^{\pm}$ into $\tau$ modes.
In the context of the fourth benchmark {\bf BP4}, light winos guarantee a
substantial trilepton signal, that is easier to detect when the presence of taus
is not required. In contrast, the spectrum of the {\bf BP5} scenario does
not allow for copious multilepton production, rendering this LRSUSY
configuration difficult to probe with electroweakino searches.

As shown by the number of events populating the {\bf G05} and {\bf H04} regions,
four-lepton final states are generally produced substantially. In most cases,
four-lepton channels, even if associated with a smaller background, cannot
compete with the trileptonic modes. Our scenario {\bf BP4} consists however
in an exception. Here, charginos mostly decay into the LSP along with electrons,
as the LSP sneutrino is of the first generation. Fewer tau leptons thus appear
in the cascade, and branchings into four light-flavored leptons are larger.

Moving on with prospects for the high-luminosity LHC, we observe in
Table~\ref{tab:nev_sig1} that the {\bf D16} region proves to be the best handle
on all LRSUSY configurations, mainly because the associated background is small.
It indeed yields almost a $5\sigma$ significance to all channels. The {\bf D16}
search region is the most significant, especially when the LSP consists in a
sneutrino of third generation, as chargino decays into taus are more common.
Not surprisingly, the {\bf BP1} and {\bf BP2} scenarios offer best discovery
prospects at high luminosity, given the large mass gaps between the neutralino
and chargino states and the sneutrino LSP, which results to hard lepton and tau
production from the electroweakino cascades. At lower luminosity for which the
{\bf D16} signal is statistically limited, other signal regions can be more
effective, like for example, the four-lepton {\bf G05} and {\bf H04}
modes that are more sensitive to the {\bf BP4} scenario featuring large mass
splittings that could yield harder leptons than the {\bf D16} region.

\begin{figure}
  \begin{center}
  \includegraphics[width=0.48\columnwidth]{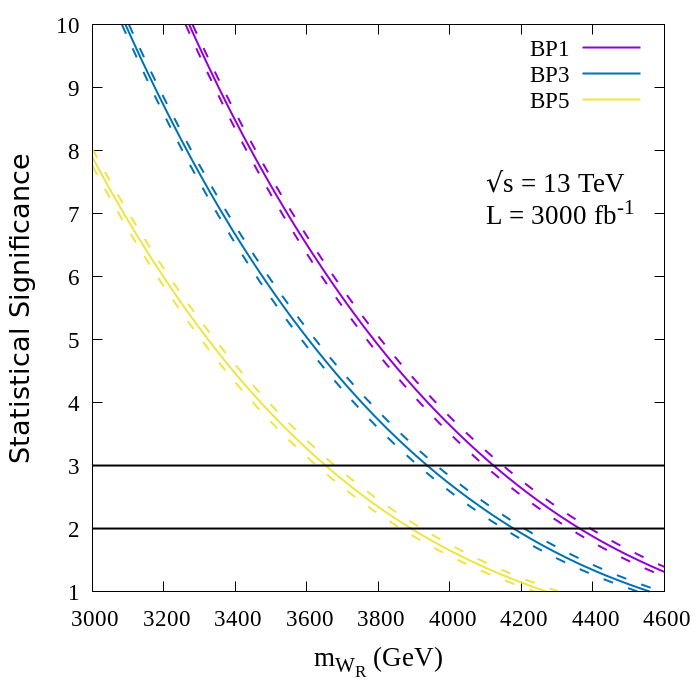}
  \includegraphics[width=0.48\columnwidth]{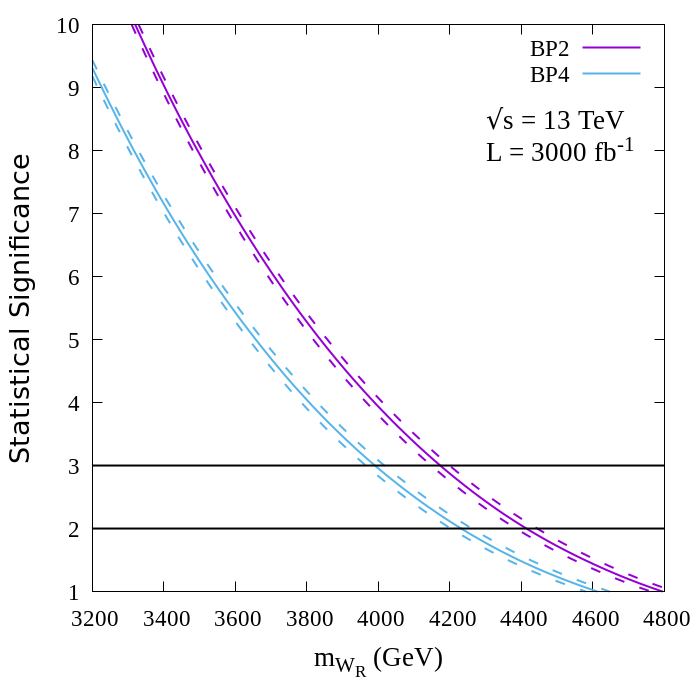}
  \end{center}
  \caption{Statistical significance of the best signal region ({\bf D16}) for
    an integrated luminosity of 3000~fb$^{-1}$ and for the five chosen benchmark
    points. The uncertainties on the background are assumed similar to the
    current ones (at 35.9~fb$^{-1}$) and we study the dependence of the results
    on the mass of the $W_R$ boson.}
  \label{fig:stat_sig}
\end{figure}

In Figure~\ref{fig:stat_sig}, we study deviations of our five benchmarks where
the mass of the $W_R$ boson is varied in the $[3, 5]$~TeV range, with all other
model parameters kept unaltered. For our predictions, we solely consider the
most sensitive search region, namely the {\bf D16} one, and focus on
3000~fb$^{-1}$ of LHC collisions at a center-of-mass energy of 13~TeV. The
impact of the $W_R$ mass on the selection efficiency is of at most $10\%$, so
that we can approximate it to be constant upon the chosen $W_R$ boson mass
range. Once again, we adopt background relative uncertainties to be equal to
those obtained at 35.9~fb$^{-1}$, which yields a conservative estimate on the
$m_{W_R}$ reach since the ratio $\sigma_B/B \simeq 0.83$ is quite large for
what concerns the {\bf D16} region. We indicate, in the figure, the
effect of a 10\% variation on the background uncertainty by dotted lines.
Once again, the {\bf BP1} and {\bf BP2} scenarios offer the best prospects, as
$W_R$ bosons as heavy as about 4.4--4.5 TeV could be excluded. The sensitivity
to the {\bf BP3} and {\bf BP4} scenarios is also quite encouraging, with an
exclusion reach of about 4.2~TeV, and the higgsino-like LSP {\bf BP5} case is
again the most complicated to probe.

One can, in principle, get an estimate of the $W_R$ boson mass from the decay
products. Whilst the final state arises from the production and decay of a
resonance, one can construct a transverse variable involving the momenta of all the
final state leptons and the missing momentum. Such an observable exhibits a
kinematic threshold that could be used to get information on the $W_R$ boson
mass. The low expected signal statistics could, however, challenge this task, but
combining varied search regions could potentially provide extra handles on the
new boson.

\section{Summary and conclusion}
\label{sec:conclusion}
We have considered minimal left-right supersymmetric new physics
scenarios and studied dark-matter-motivated configurations. In order to obtain
viable DM candidates with respect to the relic density and DM direct detection
bounds, we have demonstrated that scenarios quite compressed are in order to
guarantee a sufficient level of co-annihilations. In particular, scenarios
featuring a light sneutrinos LSP can hardly escape being ruled out when the mass
splitting between the LSP and the NLSP is large, although they are perfectly
viable when the spectrum is compressed. Heavier sneutrinos and bidoublet
higgsino LSPs are also good options, even if co-annihilation channels are here a
requirement (and automatic in the higgsino case) to ensure the agreement with
cosmological data.

We have chosen five benchmark scenarios to showcase some of the features of
these co-annihilating LRSUSY DM scenarios and investigated how they could be
probed at colliders through multileptonic signals emerging from the production
of a $W_{R}$ boson decaying into electroweakinos. Whilst we have mostly focused on
a sneutrino LSP co-annihilating with wino and higgsino setup, we have also
studied one example of higgsino LSP for comparison. For each scenario, we have
investigated the status with respect to indirect dark matter detection, focusing
on the impact of the recent results of the Fermi-LAT collaboration. We have
found that the typical DM annihilation cross sections at the present epoch lie
two orders of magnitude below the current bounds, so that all scenarios are
safe.

We have then moved on with a study of the corresponding multilepton plus missing
energy collider signals. The results are very promising, as
cosmologically-favorable configuration leads to the production of hard leptons
and taus, in association with missing energy, that could be observed through
standard electroweakino searches. We have demonstrated that by using one single of
the numerous signal regions targeting electroweakinos, the high-luminosity phase
of the LHC will allow to collect enough data to (almost) observe any of the
considered LRSUSY configurations and thus discover left-right supersymmetry
(which requires both the observation of a charged $W_R$ boson and a missing
energy signal).

\section*{Acknowledgements}
MF acknowledges NSERC for partial financial support under grant number SAP105354 and thanks the Helsinki Institute of Physics 
for their hospitality. HW acknowledges the financial support from Magnus Ehrnrooth foundation. BF has been supported in part by 
French state funds managed by the Agence Nationale de la Recherche (ANR), in the context of the LABEX ILP
(ANR-11-IDEX-0004-02, ANR-10-LABX-63). SM, KH and HW acknowledge H2020-MSCA-RICE-2014 grant no. 645722 (NonMinimal Higgs). 
AC acknowledges support from DST, India, under grant no number IFA 15 PH-130. The work of SKR was partially supported by 
funding available from the Department of Atomic Energy, Government of India, for the Regional Centre for Accelerator-based 
Particle Physics (RECAPP), Harish-Chandra Research Institute. We thank Tesla Jeltema for her comments 
on the estimation of Inverse Compton contribution, in comparison to the prompt gamma-ray flux originating from Dark Matter annihilation in the context of dSPhs.
\bibliography{lrsusy}
\end{document}